\documentclass[11pt]{article}

\usepackage{amsfonts}
\usepackage{amsmath}
\usepackage{amssymb}
\usepackage{amsthm}
\usepackage[mathscr]{eucal}
\usepackage{bbold}
\usepackage{bm}
\usepackage{braket}
\usepackage{color}
\usepackage{comment}
\usepackage{dsfont}
\usepackage{framed}
\usepackage{jheppub}
\usepackage{mathtools}
\usepackage{multirow}
\usepackage{physics}
\usepackage[normalem]{ulem}
\usepackage{subcaption}
\usepackage{tensor}
\usepackage{thmtools}
\usepackage{thm-restate}
\usepackage{enumitem}
\usepackage{pifont}
\usepackage{ulem}
\usepackage{tikz}
\usetikzlibrary{math} 
\usepackage{graphicx}
\usepackage{caption}
\usepackage{subcaption}
\usepackage{hyperref}

\makeatletter\def\@fpheader{~}\makeatother
\makeatletter
\newcommand*{\defeq}{\mathrel{\rlap{%
			\raisebox{0.3ex}{$\m@th\cdot$}}%
		\raisebox{-0.3ex}{$\m@th\cdot$}}%
	=}
\makeatother
\DeclareMathOperator{\diag}{diag}

\newcommand{\rp}{r_+}
\newcommand{\ro}{r_0}

\newcommand{\yp}{y_+}
\newcommand{\yo}{y_0}

\newcommand{\gma}{\hat{\mathcal{G}}_{\alpha}}
\newcommand{\mba}{\mathbb{a}}
\newcommand{\mbb}{\mathbb{b}}
\newcommand{\mbc}{\mathbb{c}}

\newcommand{\mbl}{\mathbb{L}_{gsb}}

\allowdisplaybreaks  

\numberwithin{equation}{section}

\title{Stability of saddles and choices of contour in the Euclidean path integral for linearized gravity: Dependence on the DeWitt Parameter}

\author{Xiaoyi Liu,$^a$} \emailAdd{xiaoyiliu@ucsb.edu}
\author{Donald Marolf,$^a$} \emailAdd{marolf@ucsb.edu}
\author{Jorge E. Santos$^b$} \emailAdd{jss55@cam.ac.uk}

\affiliation{$^a$Department of Physics, University of California, Santa Barbara, CA 93106, USA}
\affiliation{$^b$Department  of  Applied  Mathematics  and  Theoretical  Physics,  University  of  Cambridge, Wilberforce Road, Cambridge, CB3 0WA, UK}

\abstract{Due to the conformal factor problem, the definition of the Euclidean gravitational path integral requires a non-trivial choice of contour.  The present work examines a generalization of a recently proposed rule-of-thumb \cite{Marolf:2022ntb} for selecting this contour at quadratic order about a saddle. The original proposal depended on the choice of an indefinite-signature metric on the space of perturbations, which was taken to be a DeWitt metric with parameter $\alpha =-1$.  This choice was made to match previous results, but was otherwise admittedly {\it ad hoc}.  To begin to investigate the physics associated with the choice of such a metric, we now explore contours defined using analogous prescriptions for $\alpha \neq -1$.  We study such contours for
Euclidean gravity linearized about AdS-Schwarzschild black holes in reflecting cavities with thermal (canonical ensemble) boundary conditions, and we compare path-integral stability of the associated saddles with thermodynamic stability of the classical spacetimes.  While the contour generally depends on the choice of DeWitt parameter $\alpha$, the precise agreement between these two notions of stability found at $\alpha =-1$ continues to hold over the finite interval $(-2,-2/d)$, where $d$ is the dimension of the bulk spacetime.  This agreement manifestly fails for  $\alpha > -2/d$ when the DeWitt metric becomes positive definite.  However, we also find dramatic failures for $\alpha< -2$ that correlate with breakdowns of the de Donder-like gauge condition defined by $\alpha$, and at which the relevant fluctuation operator fails to be diagonalizable. This provides criteria that may be useful in predicting metrics on the space of perturbations  that give physically-useful contours in more general settings. Along the way, we also identify an interesting error in \cite{Marolf:2022ntb}, though we show this error to be harmless.}

\begin{document}

\maketitle
\section{Introduction}

In analogy with non-gravitational theories, it is generally expected that gravitational partition functions $Z(\beta)$ can be described by some notion of a Euclidean path integral \cite{Gibbons:1976ue}.  However, due to the conformal factor problem, the Euclidean action is unbounded below on the space of smooth real Euclidean metrics. As a result, the integral over the real Euclidean contour is expected to diverge.

An often-discussed potential remedy for this problem is to define the above path integral by integrating over some {\it other} contour in the space of complex metrics which gives better convergence properties.  In particular, \cite{Gibbons:1978ac} proposed that, for path integrals that compute partition functions in the canonical ensemble in asymptotically flat or asymptotically AdS spacetimes, the contour for linearized fluctuations about a saddle could be specified
by decomposing perturbations into pure-trace, transverse-traceless (TT), and pure-gauge modes.  The conformal factor problem can then be avoided by choosing to integrate the amplitudes of pure-trace modes over imaginary field values  while maintaining reality of the  TT-mode amplitudes. Since the gauge modes have vanishing action, their contour may be chosen to behave in any manner that respects the boundary conditions. Defining the path integral in this way is referred to as rotating the contour of integration for pure trace modes while leaving other modes unchanged.

However, this proposal is not self-consistent in more complicated scenarios.  In particular, \cite{Marolf:2022ntb} studied gravitational thermodynamics in a finite-sized cavity with a fixed induced-metric on the boundary.  These boundary conditions turns out to couple the pure-trace and TT modes, making it impossible to Wick rotate one and not the other.  Similar effects can occur with more familiar boundary conditions in the presence of matter, as the pure-trace modes then couple to matter fields and cannot be Wick-rotated while keeping all matter fields real.  In the context with matter, such issues have traditionally been dealt with by seeking a combination of matter and pure-trace modes that decouples, and which can thus be Wick-rotated without affecting other modes \cite{Kol:2006ga,Monteiro:2008wr,Marolf:2021kjc}.  However, this approach is not as systematic as one would like.

Such contexts thus require the prescription of \cite{Gibbons:1978ac} to be generalized.  A rule-of-thumb for this generalization was proposed in \cite{Marolf:2022ntb}, where the Wick rotation was defined by diagonalizing a specific second-order differential operator $L$ whose eigenvalues are allowed to be complex. The rule-of-thumb relies on the choice of an indefinite-signature metric on the space of metric perturbations, which was  taken to be a DeWitt metric (see \eqref{eq:DWmetric} below) with parameter $\alpha =-1$. The main justification for this choice of $\alpha$ was simply that the rule-of-thumb then coincided with the prescription of \cite{Gibbons:1978ac} in the limit where the cavity wall receded to infinity (and thus where the prescription of \cite{Gibbons:1978ac} was known to succeed).
The indefinite signature of the metric means that modes can have either positive or negative norm, and the rule-of-thumb stated that appropriate negative-norm parts of the $L$-eigenmodes should be Wick-rotated while the reality of the positive-norm parts should be preserved.  The resulting path integral then turns out to be convergent so long as the real parts of all eigenvalues of $L$ are positive. In such cases one says that the associated saddles are path-integral stable.

Using this recipe, and for boundary conditions appropriate to the canonical ensemble,  \cite{Marolf:2022ntb} found path Euclidean Schwarzschild Anti-de Sitter (ESAdS) black holes in spherical reflecting cavities to be path-integral stable saddles when the black holes have positive specific heat, and to have a single negative mode when the specific heat is negative.  Using the same recipe, the same authors showed the microcanonical Euclidean action at quadratic order to define a positive definite $L$ so that, as expected, all ESAdS black holes are stable saddles for such path integrals \cite{Marolf:2022jra}.

It is interesting to ask if the success of the rule-of-thumb is intrinsically tied to this particular choice of metric on the space of perturbations, or whether other options are equally viable.  In practice, this can be studied by attempting to use alternative metrics and examining stability of the resulting path integrals.  Some ability to change the metric is to be expected from Cauchy's theorem, which allows smooth deformations of the contour of integration within the integrand's domain of analyticity.  However, the actual extent to which this is possible remains to be understood.

Below, we investigate the simple cases obtained by varying the parameter $\alpha$ in the DeWitt$_\alpha$ metrics away from $\alpha =-1$.  As described by DeWitt in his seminal 1967 paper \cite{DeWitt:1967yk}, the DeWitt$_\alpha$ metrics form the unique ultralocal family of metrics on the space of metric perturbations $h_{ab}$ built algebraically from the background spacetime metric $g_{ab}$.  The associated line element in the space of perturbations about a $d$-dimensional spacetime $M$ with metric $\hat g_{ab}$ is given by $\int_M \gma^{abcd} h_{ab} h_{cd}$
with
\begin{equation}
\label{eq:DWmetric}
\gma^{abcd} =\frac{1}{2}(\hat{g}^{ac}\hat{g}^{bd}+\hat{g}^{ad}\hat{g}^{bc}+\alpha\hat{g}^{ab}\hat{g}^{cd}).
\end{equation}
A short calculation then shows the inverse metric to be
\begin{equation}\label{eq:InverseDeWittMetric}
    \hat{\mathcal{G}}_{\alpha\,ab\,cd}=\frac{1}{2}(\hat{g}_{ac}\hat{g}_{bd}+\hat{g}_{ad}\hat{g}_{bc}+\bar{\alpha}\hat{g}_{ab}\hat{g}_{cd}),\quad \bar{\alpha}=-\frac{2\alpha}{2+d\,\alpha},
\end{equation}
which of course satisfies
\begin{equation}
    \hat{\mathcal{G}}^{ab\,cd}_{\alpha} \hat{\mathcal{G}}_{\alpha\,ab\,ef}=\frac{1}{2}(\delta^c_e\delta^d_f
    +\delta^c_f\delta^d_e).
\end{equation}
From equation (\ref{eq:InverseDeWittMetric}), we see that $\hat{\mathcal{G}}^{ab\,cd}_\alpha$ fails to be invertible only at $\alpha=-2/d$, where the parameter $\bar \alpha$ diverges.   For $\alpha>-2/d$, the signature of $\hat{\mathcal{G}}^{ab\,cd}_\alpha$ at each spacetime point is $(0,\frac{d(d+1)}{2})$ and $\hat{\mathcal{G}}^{ab\,cd}_\alpha$ is positive-definite, so that there are no negative-norm parts for the prescription of \cite{Marolf:2022ntb} to rotate. In such cases this rule-of-thumb clearly fails as the conformal factor problem remains in full force.  We will thus be interested only in the case $\alpha<-2/d$ where the signature is indefinite, and where it is in fact $(1,\frac{d(d+1)}{2}-1)$ at each spacetime point.

In the course of our investigations, we will uncover an interesting error in \cite{Marolf:2022ntb}, as we will find that the rule-of-thumb fails to be well-defined (even for $\alpha=-1$) at what that reference called `bubble walls'.  These are codimension-1 surfaces in parameter space where the eigenvalues of an important fluctuation operator transition from being real to arising in complex-conjugate pairs.  However, detailed numerical investigation will support the claim that this is harmless at such bubble walls, as the contours on each side of the wall are nevertheless related by smooth deformations.  This is then consistent with the data from \cite{Marolf:2022ntb} showing that, at least for $\alpha =-1$, thermodynamics correctly predicts path-integral stability of all saddles on both sides of the transition.

While the value $\alpha=-1$ is not obviously distinguished at a fundamental level, it has a natural association with the familiar De Donder gauge that we will review below.  This has motivated various authors to focus on this DeWitt metric in the past for use in a variety of contexts.   In particular, it was found to define Ricci flows in which stability of ESAdS black holes was perfectly correlated with thermodynamic stability \cite{Headrick:2006ti,DeBiasio:2022nsd}.   However, the utility of other values $\alpha \neq -1$ have generally not been studied and remain to be investigated.

We begin our preparations below by using section \ref{sec:review} to review the Wick-rotation rule-of-thumb proposed in \cite{Marolf:2022ntb} and, in particular, the condition for a black hole solution to be a stable saddle of the resulting path integral.  Section \ref{sec:ModeStability} then sets the rest of the stage by reviewing the ESAdS saddles and discussing useful gauge conditions, boundary conditions, and techniques for discretizations and numerics. Section \ref{sec:NumericalResults} presents numerical results in dimension $d=4$, where we find agreement between thermodynamic stability  and stability of the Wick-rotated path integral for $\alpha\in(-2,-1/2)$. We also explicitly show that agreement fails, and in fact that it fails rather dramatically, when $\alpha\le -2$ or $\alpha>-1/2$. We identify features suggesting that there is a corresponding transition at $\alpha=-2$ in other dimensions as well.  These features also suggest a general criterion to select metrics on the space of perturbations that define physically useful contours.  Conclusions and further discussion are provided in Section \ref{sec:Conclusion}.

\section{The Rule-of-Thumb and Stability of Saddles}\label{sec:review}

This section reviews the Wick-rotation ``rule-of-thumb" recipe  proposed in \cite{Marolf:2022ntb} for perturbations of Euclidean Einstein-Hilbert gravity with action
\begin{equation}\label{eq:EHaction}
    S=-\frac{1}{16\pi G_{\mathrm{N}}}\int_{\mathcal{M}}\dd^d x\sqrt{g}(R-2\Lambda)-\frac{1}{8\pi G}\int_{\partial\mathcal{M}} \dd^{d-1}x\sqrt{\gamma}K.
\end{equation}
Generalizations to include matter and higher derivative terms are straightforward and would be interesting to explore. To organize this material for the reader, we break the discussion into 3 parts.  The first (section \ref{sec:Lops}) constructs the so-called fluctuation operators which play a fundamental role in the rule-of-thumb. The second (section \ref{sec:Wick}) reviews the Wick-rotation proposal of \cite{Marolf:2022ntb} and describes the associated notion of stability for path integral saddles.  The third (section \ref{sec:stabcomp}) reviews why it is of interest to compare path integral stability with thermodynamic stability and foreshadows an interesting puzzle to be resolved in section \ref{sec:NumericalResults}.

\subsection{Fluctuation operators for linearized gravity}
\label{sec:Lops}

Consider a perturbation $h_{ab}$ of a saddle point $\hat{g}_{ab}$.  The action for such a perturbation may be written
\begin{equation}
    S[\hat{g}+h]=S^{[0]}[\hat{g}]+S^{[2]}[h]+\text{higher order terms}.
\end{equation}
The first order term $S^{[1]}$ vanishes since $\hat{g}$ is a saddle point and thus satisfies the equation of motion.

The one-loop correction to the partition function is determined by the quadratic term $S^{[2]}[h]$.
Given any non-degenerate inner product $(h_1, h_2)_{\mathcal{\hat{G}}}$, this term can always be written in the form
\begin{equation}
\label{eq:Ldef}
    S^{[2]}[h]=(h,L h)_{\mathcal{\hat{G}}},
\end{equation}
where we take $(h_1, h_2)_{\mathcal{\hat{G}}}$ to be anti-linear in $h_1$ and linear in $h_2$, and where \eqref{eq:Ldef} uniquely determines the Hermitian linear operator $L$. This uniqueness may be seen by writing $h= \alpha_1 h_1 + \alpha_2 h_2$ for $\alpha_1, \alpha_2 \in \mathbb{C}$ for any two perturbations $h_1, h_2$.  Taking derivatives of \eqref{eq:Ldef} with respect to both $\alpha_1^*$ (where $*$ denotes complex conjugation) and $\alpha_2$ (while holding fixed $\alpha_1$ and $\alpha_2^*$) yields
\begin{equation}\label{eq:Qaction_general2}
   \frac{\partial^2}{\partial \alpha^*_1 \partial \alpha_2 } S^{[2]}[\alpha_1 h_1 + \alpha_2 h_2]= (h_1,L h_2)_{\mathcal{\hat{G}}},
\end{equation}
from which one may compute all matrix elements of $L$. In particular, for all $h_1,h_2$ we have
\begin{eqnarray}\label{eq:Qaction_general3}
(h_1,L h_2)_{\mathcal{\hat{G}}} &=&   \frac{\partial^2}{\partial \alpha^*_1 \partial \alpha_2 } S^{[2]} \cr &=&  \left[ \frac{\partial^2}{\partial \alpha^*_2 \partial \alpha_1 } S^{[2]} \right]^*  = (h_2,L h_1)^*_{\mathcal{\hat{G}}} = (L h_1, h_2)_{\mathcal{\hat{G}}},
\end{eqnarray}
so that $L$ is Hermitian as claimed.  For lack of a better name, we will refer to $L$ as the {\it fluctuation operator} below.

Recall that the fluctuation operator $L$ is defined by the choice of $\mathcal{\hat{G}}$.  As discussed in \cite{Marolf:2021kjc}, an interesting class of such  inner products is given by the DeWitt$_\alpha$ metrics  ${\mathcal{G}}^{ab\,cd}_\alpha$ from \eqref{eq:DWmetric} with
\begin{equation}
\label{eq:DWIP}
    (h,\tilde{h})_{\hat{\mathcal{G}}_\alpha}=\frac{1}{32\pi G}\int_{\mathcal{M}}\dd ^d x \,h_{ab}^*\hat{\mathcal{G}}^{ab\,cd}_\alpha\tilde{h}_{cd},
\end{equation}
where $*$ denotes complex conjugation.  We will use $L_\alpha$ to denote the fluctuation operator determined by the inner product \eqref{eq:DWIP}. This $L_\alpha$ is
is a second-order linear differential operator whose details depend on both the chosen saddle and the value of $\alpha$.

The case $\alpha=-1$ was studied in \cite{Marolf:2022jra} for fluctuations $h_{ab}$ inside a cavity that leave the induced (Euclidean) spacetime metric unchanged on the cavity walls.  In this context, the fluctuation operator was found to be
\begin{equation}\label{eq:lich}
    \begin{split}
        (L_{-1}h)_{ab}&=(\hat{\Delta}_L h)_{ab}+2\hat{\nabla}_{(a}\hat{\nabla}^p\bar{h}_{b)p},\\
        \bar{h}_{ab}=h_{ab}-\frac{\hat{g}_{ab}}{2}h,&\quad (\hat{\Delta}_L h)_{ab}=-\hat{\nabla}_p\hat{\nabla}^ph_{ab}-2\hat{R}_{acbd}h^{cd},
    \end{split}
\end{equation}
in terms of the standard Lichnerowicz operator $\hat{\Delta}_L$.

Below, we will be interested in studying the same linearized action $S^{[2]}[h]$ while taking the inner product to be given by a general DeWitt${}_\alpha$ metric \eqref{eq:DWmetric}.
Since the DeWitt$_\alpha$ metrics are ultralocal, we must have
\begin{equation}
    \hat{\mathcal{G}}^{abcd}_{\alpha}(L_{\alpha}h)_{cd}=\hat{\mathcal{G}}^{abcd}_{-1}(L_{-1}h)_{cd}.
\end{equation}
Using the inverse DeWitt$_\alpha$ metric $\hat{\mathcal{G}}_{\alpha\,abef}$ from \eqref{eq:InverseDeWittMetric}, we then find
\begin{equation}
\label{eq:Lalpha}
 (L_{\alpha}h)_{ef}=\hat{\mathcal{G}}_{\alpha\,abef}\hat{\mathcal{G}}^{abcd}_{-1}(L_{-1}h)_{cd}
=(L_{-1}h)_{ef}-\frac{1+\alpha}{2+d\alpha}\Tr(L_{-1}h)g_{ef},
\end{equation}
where $\Tr(L_{-1}h)=\hat{g}^{ab}(L_{-1}h)_{ab}$. As expected from the use of the inverse DeWitt$_\alpha$ metric, this $L_\alpha$ is well-defined for $\alpha \neq -2/d$ (though it diverges for $\alpha =-2/d$).
Taking the trace of \eqref{eq:Lalpha} yields the relation
\begin{equation}
\label{eq:LalphaTr}
 \Tr (L_{-1}h)
= \frac{2+d\alpha}{2 -d}\Tr(L_{\alpha}h).
\end{equation}

For future use, we note that comparing \eqref{eq:Lalpha} with \eqref{eq:LalphaTr} shows that any $h$ annihilated by $L_{-1}$ is also annihilated by all $L_\alpha$ for all $\alpha \neq -2/d$, and also that for $d>2$ any $h$ annihilated by any well-defined $L_\alpha$ is annihilated by $L_{-1}$.  In other words, the zero-eigenvalue eigenvectors of our $L_\alpha$ are manifestly independent of $\alpha$.

\subsection{The Wick rotation rule-of-thumb}
\label{sec:Wick}

The proposal of \cite{Marolf:2022ntb} was to select a Wick-rotation by first choosing a non-degenerate inner product $\hat{\mathcal {G}}$ and constructing the associated fluctuation operator $L$.  As noted above, this $L$ is Hermitian.  It is then natural to assume that $L$ can be diagonalized.  Indeed, the rule-of-thumb supposes that this is so,  and also that the spectrum is non-degenerate (say, up to manifest symmetries of the system).  When either of these conditions fail, the rule-of-thumb is generally not well-defined.  While it will generally be possible to deform the setting slightly so as to make the perturbed $L$ diagonalizeable, we will see below that there are interesting cases in which the associated Wick  rotations fail to have a well-defined limit when this perturbation in removed.

It is important to note that failures of $L$ to be diagonalizable can result from two distinct sorts of issues.  The first is that $L$ is always Hermitian, but in an infinite-dimensonal inner product space this does not necessarily imply that $L$ is in fact self-adjoint; see e.g. \cite{RS} for discussions of this point in the context of a positive-definite inner product.  However, for second-order differential operators, one generally finds that $L$ is self-adjoint if one has taken sufficient care in dealing with boundary conditions.  This is deeply related to the fact that we expect to be able to approximate our continuous system with a discretization in which the associated inner product space is of finite dimension.  In particular, we can take the discretized system to be defined by a discretized quadratic action $\mathbb{S}^{[2]}$. If we then define a discretized fluctuation operator $\mathbb{L}$ via the analog of \eqref{eq:Qaction_general2}, the result \eqref{eq:Qaction_general3} will again show  $\mathbb{L}$ to be Hermitian.  But,  in a finite-dimensional inner product space, any Hermitian $\mathbb{L}$ is in fact self-adjoint, arguing that $L$ can at least be approximated by a self-adjoint discretized operator.

The second issue, however, is simply that our inner product $(,)_{\hat{\mathcal{G}}}$ has indefinite signature.  In this case there are always self-adjoint operators that cannot be diagonalized.  A well-known finite-dimensional example involves the operator
\begin{equation}
    \mathcal{O}=\begin{pmatrix}
2 & -1 \\
1 & 0
\end{pmatrix}
\end{equation}
on the two-dimensional inner product space with the Minkowski metric
 \begin{equation}
        \hat{\mathcal{G}}=\diag(1,-1).
    \end{equation}
One may readily check that $\mathcal{O}^\dagger \hat{\mathcal{G}}= \hat{\mathcal{G}} \mathcal{O}$, where $^\dagger$ denotes the complex-conjugate transpose, and that this condition is equivalent to self-adjointness of $\mathcal{O}$ with respect to the inner product $(,)_{\hat{\mathcal{G}}}$. However, $\mathcal{O}$ is not diagonalizable since the characteristic polynomial is $(\lambda-1)^2$ (so that the `algebraic multiplicity' is $2$ for the eigenvalue $\lambda = 1$) while the only  eigenvector is
$\left[ \begin{smallmatrix}
  1\\1
\end{smallmatrix}\right]$
 (so that the `geometric multiplicity' of that eigenvalue is only $1$).  On the other hand,  if the characteristic polynomial $\det \left({\cal O} - \lambda \mathbb{1} \right)$ of any $d \times d$ matrix $\mathcal{O}$ has $d$ distinct roots, then $\mathcal{O}$ can always be diagonalized.  As a result, at least in finite-dimensional inner product spaces, we may say that diagonalizable operators are generic, and that non-diagonalizable operators only arise at a measure-zero set of parameters.   However, non-diagonalizability can still lead to important phenomena when we continuously  deform our operator $L$ by varying $\alpha$.

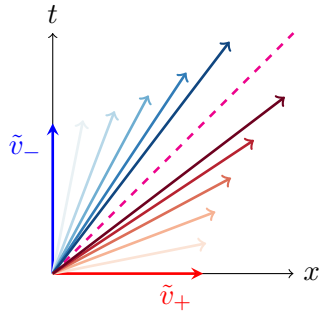
\begin{figure}[h!]
\centering
\begin{tikzpicture}[scale=2]
\definecolor{red1}{RGB}{251,227,213};
\definecolor{red2}{RGB}{246,178,147};
\definecolor{red3}{RGB}{220,109,087};
\definecolor{red4}{RGB}{183,034,048};
\definecolor{red5}{RGB}{109,001,031};
\definecolor{blue1}{RGB}{233,241,244};
\definecolor{blue2}{RGB}{182,215,232};
\definecolor{blue3}{RGB}{109,173,209};
\definecolor{blue4}{RGB}{049,124,183};
\definecolor{blue5}{RGB}{016,070,128};
  \draw[->] (0,0)--(1.6,0) node[right]{$x$};
  \draw[->] (0,0)--(0,1.6) node[above]{$t$};
  \draw[line width=1pt,blue,-stealth](0,0)--(0,1) node[anchor=north east]{$\tilde v_-$};
  \draw[line width=1pt,red,-stealth](0,0)--(1,0) node[anchor=north east]{$\tilde v_+$};
  \draw[line width=1pt, magenta,dashed](0,0)--(1.6,1.6);
\draw[->,blue1,line width=1pt] (0,0)--(0.201336,1.02007);
\draw[->,blue2,line width=1pt] (0,0)--(0.410752,1.08107);
\draw[->,blue3,line width=1pt] (0,0)--(0.636654,1.18547);
\draw[->,blue4,line width=1pt] (0,0)--(0.888106,1.33743);
\draw[->,blue5,line width=1pt] (0,0)--(1.17520,1.54308);
\draw[->,red1,line width=1pt] (0,0)--(1.02007,0.201336);
\draw[->,red2,line width=1pt] (0,0)--(1.08107,0.410752);
\draw[->,red3,line width=1pt] (0,0)--(1.18547,0.636654);
\draw[->,red4,line width=1pt] (0,0)--(1.33743,0.888106);
\draw[->,red5,line width=1pt] (0,0)--(1.54308,1.17520);
\end{tikzpicture}
\caption{Orthogonal pairs of negative- and positive-norm vectors are shown in a fixed 1+1 Minkowski space.  The vectors are normalized with respect to a $1+1$ Minkowski metric, so the vectors diverge in the limit where they become null.   In cases of interest the metric will turn out to vary at the same order at which the vectors fail to be null, and thus at which the two eigenvalues differ.  As a result, this picture is {\it not} an accurate depiction of the general case, though normalized vectors will always diverge in the null limit.} \label{fig:approachnull}
\end{figure}

In particular, suppose that a postive-norm mode and a negative-norm mode become degenerate with common eigenvalue $\lambda$ at some value $p_*$ of the parameters (while all other modes maintain distinct eigenvalues).  Suppose also that this happens in such a way that norm of each mode vanishes when the degeneracy occurs.
Note that near $p_*$ we may consider the modes to live in a fixed two-dimensional space defined by being orthogonal to all other modes.   The inner product on our two-dimensional space will typically vary with the parameters $p$, either because $p$ enters implicitly in the background metric which defines some DeWitt metric $\gma$ or because we explicitly include the DeWitt parameter $\alpha$ in the parameters $p$. However,  this variation is small over small ranges of $p$ near $p_*$, so that each $p$ will define a pair of eigenvectors that is both nearly-orthogonal and nearly-null.  As a result,
each eigenvector must approach the same null vector $v_1$ in the limit $p \rightarrow p_*$, so that we obtain only a single eigenvector in this limit; see figure \ref{fig:approachnull} for the familiar example of exactly orthogonal vectors becoming null in the limit of a large boost in a fixed 1+1 Minkowski metric.  If $v_2$ is the conjugate null vector (with $(v_1, v_2) = 1$) then self-adjointness requires $\lambda  = \lambda(v_2, v_1) = (v_2, L v_1) = (Lv_2,  v_1)$.  But since $v_1$ is null, this allows
\begin{equation}
\label{eq:gamma}L v_2 = \lambda v_2 + \gamma v_1
\end{equation}
for any $\gamma$.  Generically one finds $\gamma \neq 0$ and $L$ fails to be diagonalizable at $p_*$.

In any case, the rule-of-thumb is defined when $L$ can be diagonalized.  Here we should also note that, as seen in the above argument,  the spectrum of $L$ can admit complex eigenvalues so long as they arise in complex-conjugate pairs.  In particular, if $\lambda$ is an eigenvalue with eigenvector $v$, then (since a real action will lead to real $L$) its complex conjugate $\lambda^*$ is an eigenvalue whose associated eigenvector $v^*$ is the complex conjugate of $v$.  Furthermore, both vectors must have vanishing norm $(v,v)_{\hat{\mathcal{G}}} = (v^*,v^*)_{\hat{\mathcal{G}}}=0$.

Unless there are degeneracies in the spectrum, the above discussion (involving $v_1, v_2)$ then shows that $(v^*, v)_{\hat{\mathcal{G}}}$ must be non-zero for eigenvectors $v\neq 0$ with complex eigenvalues.
A natural normalization condition for the eigenvectors is thus to impose
 \begin{equation}
 \label{eq:vnorm}
(v,v^*)_{\hat{\mathcal{G}}}=(v^*,v)_{\hat{\mathcal{G}}}=1.
 \end{equation}
 Note that this condition involves the complex conjugate $v^*$, while the norm of a vector is simply $(v,v)_{\hat{\mathcal{G}}}$.  As a result,  the condition \eqref{eq:vnorm} provides a convention to fix the overall phase of the vector as well as its magnitude.   With this convention, if an eigenvector $v$ is associated with a complex eigenvalue\footnote{We use the term complex eigenvalue to we refer to those whose imaginary part is non-zero.  In particular, this includes the case where $\lambda$ is purely imaginary.   The corresponding eigenvectors $v$ are intrinsically complex (since $v$ and $v^*$ have different eigenvalues), and we refer to them as complex eigenvectors below.  In contrast, when $\lambda$ is real the corresponding eigenvectors are real up to an overall phase.  In a slight abuse of terminology, we refer to these as real eigenvectors below regardless of how the phase is chosen, and in particular regardless of whether or not \eqref{eq:vnorm} is consistent with setting $v=v^*$.}
 then, since $(v,v)_{\hat{\mathcal{G}}} =(v^*,v^*)_{\hat{\mathcal{G}}}=0$ as described above, the condition \eqref{eq:vnorm} requires the real part $\Re v=(v+v^*)/2$ of the eigenvector to be orthogonal to the imaginary part $\Im v=(v-v^*)/2i$.  We also then find the real part to have positive norm, while the norm of the imaginary part is necessarily negative.
The eigenvectors $v_\lambda$ can thus be divided into two categories:
\begin{enumerate}
    \item Real eigenvectors that have a positive norm, and the real parts of the complex eigenvectors.
    \item Real eigenvectors that have a negative norm, and the imaginary parts of the complex eigenvectors.
\end{enumerate}

The proposal of \cite{Marolf:2022ntb} is simply that, when  ${L}$ is diagonalizable and its spectrum is nondegenerate, perturbations corresponding to vectors in category $2$ should be Wick-rotated by multiplying them by a factor of $i$, while vectors in category $1$ are left invariant.
A short computation then shows that any mode with  $\Re \lambda>0$ has positive action after the above Wick rotation.  We will therefore say that modes with $\Re \lambda>0$ are stable modes under this proposal (regardless of the sign of their norm).

Due to gauge invariance, the spectrum of the fluctuation operator $L$ defined above will in fact be highly degenerate at eigenvalue $\lambda =0$. However, the rule-of-thumb can remain well-defined in this case since one finds identical results whether or not one chooses to Wick-rotate any pure-gauge mode.  This occurs because pure-gauge modes always have vanishing action, so that their contribution to the action still vanishes even if the mode has been multiplied by a factor of $i = \sqrt{-1}$.

Nevertheless,  it will be useful in practice to break the gauge invariance by adding an \textit{ad hoc} term to the quadratic action $S^{[2]}$.  The operator $L_{\text{gsb}}$ we will use in our numerics below will again be defined by \eqref{eq:Ldef}, but where we now use the modified gauge-symmetry-broken action instead of the gauge-invariant one.  The treatment of gauge modes will be discussed further in section \ref{sec:gaugeCondition}.

\subsection{Path Integral vs thermodynamic stability}
\label{sec:stabcomp}

The purpose of this work is to examine the viability of the rule-of-thumb contour prescription of \cite{Marolf:2022ntb} for various choices of metric on the space of perturbations.  Since the fundamental prescription that determines this contour is not currently understood, the main tool at hand for testing this viability is compatibility with the semiclassical physics, and with what we expect the semiclassical physics to imply.
As in  \cite{Prestidge:1999uq,Marolf:2022ntb}, we will focus on the fact that Hawking's assignment of a temperature to each stationary black hole \cite{Hawking:1975vcx} allows us to compute a notion of specific heat, and on the expectation that Euclidean path integral computes a useful notion of a partition function $Z = \Tr \ e^{-\beta H}$ for our theory of quantum gravity.

To remind the reader of the implications of these points, it is useful to recall some basic facts from quantum statistical mechanics.  Suppose therefore that we are given a standard quantum system, with positive definite Hamiltonian operator, and that we wish to computer the standard partition function.
If we can approximate the sum over states by an integral with density of states $e^{{\cal S}(E)}$,
\begin{equation}
\label{eq:Zstat}
Z(\beta) : = \Tr \ e^{-\beta H}  \approx \int \dd E\, e^{{\cal S}(E)} e^{-\beta E}\equiv \int \dd E\, e^{-\beta F},
\end{equation}
and if we take a limit in which ${\cal S}-\beta E$ becomes large, then at leading order our $Z(\beta)$ is just $e^{{\cal S}(\hat E)-\beta \hat E}$ for the energy $\hat E$ that minimizes the quantity $\beta F : = \beta E - {\cal S}(E)$.  Further improvements to this approximation can then be generated by expanding ${-\beta F}$ perturbatively about $\hat E$,  keeping the quadratic term in the exponential while expanding the exponential of higher terms.  At any given order in the latter expansion, the computation reduces to integrating a polynomial times a Gaussian, which can be done in closed form.  In particular, the associated integrals always converge.

Recall also that, as a result of minimizing $\beta F$, the specific heat at $\hat E$ is necessarily positive. Indeed, if we instead choose a local maximum $\tilde E$ of $\beta F$ and attempt to perform the above perturbative expansion of $e^{-\beta F}$, then we of course find that the supposed Gaussian has negative variance (i.e., the exponent has the wrong sign) so that the perturbative integrals fail to converge.  This is equivalent to the statement that,  if we formally define a specific heat at $\tilde E$ using the free energy $F: = (\beta E -{\cal S}(E))/\beta$, then the resulting specific heat will be {\it negative}.

Let us then conjecture i) that the path integral should compute something like \eqref{eq:Zstat}, ii) that we may take the Euclidean action $S$ to model the product $\beta F$, and iii) that the semiclassical black hole specific heat reflects the statistical mechanics of an underlying quantum system.  In this case we find that saddles (classical Euclidean solutions) with positive semiclassical specific heat should be local minima of the Euclidean action $S$  along the contour of integration, while saddles with negative semiclassical specific heat should be local maxima.  In this work, we follow the standard convention of referring to local minima as `stable saddles' of the path integral and referring to local maxima as `unstable saddles.'  It is also useful to use the term `marginally stable saddle' when the second derivative of the Euclidean action $S$ vanishes along the contour.
In the multi-dimensional context we say that a saddle is path-integral-unstable when the second derivative is negative in at least one direction along the contour.  Similarly, the saddle is marginally stable when some such second derivative vanishes but no second derivative is negative.
A good contour prescription should thus have the property that thermodynamic stability of saddles (defined by the sign of the specific heat) agrees in all cases with this notion of path integral stability of saddles.

It is this agreement that we test below, and which \cite{Marolf:2022ntb} found to hold for all Euclidean Schwarzschild AdS black holes in spherical reflecting cavities using the contour ${\cal C}_{-1}$ defined by the DeWitt$_{-1}$ metric.  Here we will study alternative contours ${\cal C}_\alpha$ that are again defined using the recipe of section \ref{sec:Wick}, but which are now based on the DeWitt$_\alpha$ metrics for any $\alpha \neq -2/d$.  We may thus expect the family of test-contours  ${\cal C}_\alpha$ to be continuous,  and that at a given saddle $\hat s$ the second derivative of the Euclidean action along the contour again changes continuously with $\alpha$. In particular, a given saddle $\hat s$ would change from being path-integral-stable to being path-integral-unstable if and only if we passed through a contour on which $\hat s$ was {\it marginally stable} in the path integral sense, which in particular would require that for some ${\cal C}_\alpha$ there is a direction along the contour for which the second derivative of the Euclidean action $S$ vanishes at $\hat s$.

Since thermodynamics and path integral stability agree for the contour ${\cal C}_{-1}$, if we wish to study the success or failure of this agreement as we deform $\alpha$ away from $-1$,  it is sufficient under the above assumptions to check whether the set of saddles that are marginally stable thermodynamically coincides with the set of saddles that are marginally stable in the path integral sense (and whether in each case stability is marginal to the same order).  If this is always the case, then the agreement between the two notions of stability found at ${\cal C}_{-1}$ would be maintained for all $\alpha$.  But failure of the marginal stability test at some $\alpha$ generally implies that agreement between the two notions of stability has ceased to hold.

The important point here is that, on the Wick-rotated contour defined by the recipe of section \ref{sec:Wick}, the inner product $(,)_{\hat{\mathcal{G}}}$ becomes positive definite for directions along the contour.  If $L$ is diagonalizable, then  (by \eqref{eq:Ldef}) it describes second derivatives of the Euclidean action.  Vanishing of a second derivative of $S$ along the Wick-rotated contour is thus equivalent to finding an eigenvector of $L$ with eigenvalue zero.  However, at the end of section \ref{sec:Lops} we found that zero-eigenvalues of the fluctuation operators $L_\alpha$ coincide with those of $L_{-1}$ for all $\alpha$ in the allowed range ($\alpha < -2/d$).  This may then seem to imply that we have nothing left to check, and that thermodynamic and path integral stability must agree for all $\alpha$.  Nevertheless, we will see in \ref{sec:NumericalResults} that this is {\it not} the case!  It will then be interesting to identify precisely which of the above assumptions fails to hold.

\section{Setting the stage}
\label{sec:ModeStability}

Having reviewed the general rule-of-thumb for Wick rotations described in \cite{Marolf:2022ntb}, we now turn to the details of the setting to be explored below.  The goal of this section is to set the stage for the presentation of results in section \ref{sec:NumericalResults}.  To do so, we review ESAdS black holes in reflecting cavities that define the saddles to be studied, describe our treatment of gauge issues in linearized gravity, examine details of the boundary conditions to be imposed, and explain the discretization and numerical methods to be employed.  These items are addressed one-by-one in the subsections below.

\subsection{Saddles and Thermodynamic Stability}
Following \cite{Marolf:2022ntb}, we focus on Euclidean Schwarzschild AdS (ESAdS) black holes at the center of spherical reflecting cavities.  We also impose boundary conditions on the cavity walls that fix the induced metric.  We work in Schwarzschild coordinates and use $\rp$ and $\ro$ to respectively denote the horizon radius and the location of the cavity wall.

In $d$ spacetime dimensions the  ESAdS metric $\hat g$ is given by
\begin{equation}
    \widehat{\dd s}^2=f(r)\dd\tau^2+f^{-1}(r)\dd r^2+r^2 \dd \Omega_{d-2}^2,\quad \tau\sim\tau+\beta_*,
\end{equation}
where
\begin{equation}
    f(r)=\frac{r^2}{\ell^2}+1-\left(\frac{\rp}{r}\right)^{d-3}\left(\frac{\rp^2}{\ell^2}+1\right),\quad d\ge 3
\end{equation}
is the blackening factor, $\dd\Omega_{d}^2$ is the metric on the unit $d$-sphere, $\beta_*=4\pi \sqrt{f(\ro)}/|f'(\rp)|$ is the period of $\tau$ to avoid conical singularity, and $\ell$ is the AdS scale given by
\begin{equation}
    \ell^2=-\frac{(d-1)(d-2)}{2\Lambda},
\end{equation}
where $\Lambda$ is the cosmological constant. A short computation shows there to be two values of $\rp$ for each fixed $\beta_*$.   The larger one is well known to be thermodynamically stable while the smaller is not.

By calculating the specific heat, we can find the boundary which divides the large and small black hole solutions for a fixed cavity size:
\begin{equation}\label{eq:ypstar}
\begin{split}
    y_+^*(\yo)=\Bigg\{&\frac{d-3}{2(d-1)}-\frac{3d-11+2\yo^{d-3}}{2(d-3+2\yo^{d-1})}+\\
    &\sqrt{\left[\frac{d-3}{2(d-1)}-\frac{3d-11+2\yo^{d-3}}{2(d-3+2\yo^{d-1})}\right]^2-\frac{d-3}{(d-1)\yo^2}\left[\frac{d-3+(d-1)\yo^2}{d-3+2\yo^{d-1}}-1\right]}\Bigg\}^{1/2},
    \end{split}
\end{equation}
where
\begin{equation}
    \yo\equiv \frac{r_0}{\rp}\quad \text{and}\quad \yp\equiv \frac{\rp}{\ell}
\end{equation}
are two dimensionless quantities we will be using in presenting our numerical results. Note that $\yp$ vanishes only in the asymptotically flat case $\Lambda=0$, .

\subsection{Gauge Condition}\label{sec:gaugeCondition}

Due to the gauge symmetry of gravitational systems, any fluctuation operator $L$ defined by \eqref{eq:Ldef} will be highly degenerate at eigenvalue $\lambda=0$.  In particular, since the linearized action $S^{[2]}(h)$ is invariant under the addition $h_{ab}  \rightarrow h_{ab} + \hat{\nabla}_{(a}\xi_{b)}$ for appropriate vector fields $\xi^a$, taking $(h_2)_{ ab} =\hat{\nabla}_{(a}\xi_{b)}$ in \eqref{eq:Qaction_general2} and using the non-degeneracy of the inner product shows that $L$ must annihilate any pure-gauge mode $\hat{\nabla}_{(a}\xi_{b)}$.   In our analysis below, we will take the space of gauge transformations to be described by vector fields $\xi^a$ that vanish at the cavity wall; i.e., for which $\xi^a|_{r=r_0}=0$. Boundary conditions will be discussed in detail in the following section.

The above-mentioned degeneracies complicate the numerical computations we wish to perform, especially due to our focus on eigenvectors with eigenvalue $\lambda=0$ as discussed in section \ref{sec:stabcomp}. We can avoid such complications by choosing a particular gauge in which to study the system.  As in \cite{Marolf:2021kjc}, we will then modify the action in a manner that leaves the action unchanged for perturbations satisfying the gauge condition, but which {\it does} change the action for pure-gauge perturbations so as to give non-zero eigenvalues to the pure-gauge modes.  Note that, by definition,  pure-gauge perturbations will never satisfy a condition that completely fixes the gauge.

Given a metric on the space of perturbations, it is natural to attempt to choose a gauge condition that is satisfied precisely by the space $W^\perp$ of perturbations that are orthogonal to the space $W$ spanned by pure-gauge modes.  We take the pure-gauge modes to be of the form $h_{ab} =\tilde h_{ab}=\hat{\nabla}_{(a}\xi_{b)}$ with $\xi^a|_{r=r_0}=0$.  If we use $\mathfrak{G}$ to denote the space of all perturbations $h$ that satisfy the given boundary conditions, then we can find such a condition when any $h\in \mathfrak{G}$ can be written in the form $h= h^\perp + h^W$ for $h^W \in W$ and a unique $h^\perp \in W^\perp$.  Note that the uniqueness condition requires $W^\perp \cap W = \{0\}$, which says that there can be no non-zero $h^W \in W$ that is orthogonal to the entire space $W$.  In other words, the uniqueness condition holds precisely when the induced metric on $W$ is non-degenerate (as defined by the inner product $(,)_{\hat{\mathcal{G}}}$  on $\mathfrak{G}$).  Non-degeneracy of the induced metric on $W$ can also be used to show existence of the decomposition  $h= h^\perp + h^W$ by introducing a basis on $W$ and explicitly constructing the associated projection from $\mathfrak{G}$ to $W^\perp$. Thus we see that this construction defines a good gauge fixing precisely when the induced metric on $W$ is non-degenerate.

For later purposes, it will be useful to restate the above prescription in slightly different terms.  We begin by noting that the construction of the pure-gauge mode $h_{ab} =\tilde h_{ab}=\hat{\nabla}_{(a}\xi_{b)}$ from the vector field $\xi^b$ can be described by a linear map ${\cal P}: V \rightarrow W \subset \mathfrak{G}$, where $V$ where is the space of smooth vector fields with $\xi^a|_{r=r_0}=0$, and where $({\cal P}\xi)_{ab} = \hat{\nabla}_{(a}\xi_{b)}$.  Furthermore, by definition a mode $h$ lies in $W^\perp$ if the inner product $(h, {\cal P} \xi)_{\hat{\mathcal{G}}}$ vanishes for all $\xi \in V$.  If we then introduce a positive-definite Hermitian inner product $(\xi_1, \xi_2)_V$ on $V$, we can attempt to define an adjoint operator
$\mathcal{P}^\dagger: \mathfrak{G}\to V$ by the relation
\begin{equation}
\label{eq:Padjoint}
    (h,\mathcal{P}\xi)_{\gma}=(\mathcal{P}^\dagger h,\xi)_{V}.
\end{equation}
If this attempt succeeds, we may then express the statement that $h \in W^\perp$ as the gauge-fixing condition
\begin{equation}
\label{eq:newdedonder}
    \mathcal{P}^\dagger h=0.
\end{equation}
Note that the definition of $\mathcal{P}^\dagger$ depends both on the choice inner product on $\mathfrak{G}$ and the choice of inner product $V$.

If we take $(,)_{\hat{\mathcal{G}}}$ to be the DeWitt$_\alpha$ inner product, we may write
\begin{equation}
\label{eq:alphagforth}
32\pi G(h,\mathcal{P} \xi)_{\hat{\mathcal{G}}}=\int_{\mathcal{M}}\dd^d x\sqrt{\hat{g}}(2h_{ab}\hat{\nabla}^a\xi^b+\alpha h\hat{\nabla}_a\xi^a)=-2\int_{\mathcal{M}}\dd^d x\sqrt{\hat{g}}\left(\hat{\nabla}^a h_{ab}+\frac{1}{2}\alpha\hat{\nabla}_b h\right)\xi^b.
\end{equation}
Choosing the inner product on $V$ to be
\begin{equation}
    (\xi^a,\tilde{\xi}^b)_{V}=\frac{1}{32\pi G}\int_{\mathcal{M}}\dd^dx\ \sqrt{\hat{g}}\cdot\hat{g}_{ab}\xi^a\tilde{\xi}^b,
    \end{equation}
we can then easily read off from \eqref{eq:alphagforth} and \eqref{eq:Padjoint} that we have
\begin{equation}
\label{eq:Pdagonh}
    (\mathcal{P}^\dagger h)_a=- \left(2\nabla^bh_{ab}+\alpha\nabla_a h\right).
\end{equation}

Since we require all $\xi \in V$ to vanish at $r=r_0$, the result \eqref{eq:Pdagonh} lies in $V$ only when we restrict $h$ so that \eqref{eq:Pdagonh} vanishes at $r=r_0$.  We thus impose this boundary condition below on all $h \in \mathfrak{G}$.  This introduces a small $\alpha$-dependence in the definition of $\mathfrak{G}$ which may appear to complicate various arguments.  Note, however, we could instead choose to use $\alpha$-independent space $\mathfrak{G}$ and to take limit where the metric on $V$ becomes small at $r_0$, so that ${\cal P}^\dagger$ becomes correspondingly large at $r=r_0$.  This then pushes to large eigenvalues any mode which fails to satisfy the gauge condition at $r=r_0$.  One can then check numerically that, in this limit, all eigenmodes of $L$ satisfy the desired gauge condition.  We will thus generally ignore the $\alpha$-dependence of $\mathfrak{G}$  with the idea that arguments can be reformulated as just described. This idea is then supported by our numerical results, which will be seen to be in precise accord with the resulting expectations.

For $\alpha=-1$ we see that  \eqref{eq:newdedonder} is just the well-known de Donder gauge condition.  However, for more general $\alpha$ \eqref{eq:newdedonder} imposes the de Donder-like gauge
\begin{equation}
   \hat{\nabla}^a h_{ab}+\frac{1}{2}\alpha\hat{\nabla}_b h=0.
\end{equation}

As noted above, our gauge fixing will be incomplete if (and only if) the space $W$ of pure-gauge modes has non-trivial intersection with $W_\perp$; i.e., iff
the condition \eqref{eq:newdedonder} holds for some $h = \mathcal{P} \xi$, so that
\begin{equation}
\label{eq:PdagP}
\mathcal{P}^\dagger \mathcal{P} \xi=0
\end{equation}
for some non-trivial $\xi \in W$.  In other words, the gauge fixing  \eqref{eq:newdedonder} is complete if and only if the operator $G := \mathcal{P}^\dagger \mathcal{P}$ is an invertible map from $V$ to itself.

For the case $\alpha=-1$, there is a well-known analytic argument that $G$ is invertible for pure Einstein-Hilbert gravity (without matter) with vanishing or negative cosmological constant; see e.g. appendix A of \cite{Marolf:2022ntb}.  This argument can be generalized to study general values of $\alpha$.  To do so, note that \eqref{eq:PdagP} takes the explicit form
\begin{eqnarray}\label{eq:nondegeneracyOfG}
(G\xi)_b &=& -\left[\hat{\nabla}^2{\xi}_b+\hat{\nabla}^a\hat{\nabla}_b{\xi}_a+\alpha\hat{\nabla}_b\hat{\nabla}^a{\xi}_a \right]\cr
&=&- \left[\hat{\nabla}^2{\xi}_b+ 2\hat{\nabla}_{[a}\hat{\nabla}_{b]}{\xi}^a+ (1+\alpha) \hat{\nabla}_b\hat{\nabla}_a{\xi}^a \right]\cr
&=&-\left[\hat{\nabla}^2{\xi}_b -R_{cba}{}^c{\xi}^a+ (1+\alpha) \hat{\nabla}_b\hat{\nabla}_a{\xi}^a \right]\cr
&=&-\left[\hat{\nabla}^2{\xi}_b + R_{ba}{\xi}^a+ (1+\alpha) \hat{\nabla}_b\hat{\nabla}_a{\xi}^a \right]\cr
&=&-\left[\hat{\nabla}^2{\xi}_b + \frac{2\Lambda}{d-2}{\xi}_b+ (1+\alpha) \hat{\nabla}_b\hat{\nabla}_a{\xi}^a \right].
\end{eqnarray}
Contracting \eqref{eq:nondegeneracyOfG} with $\sqrt{\hat g} \xi^b$ and integrating over the spacetime $M$ yields
\begin{eqnarray}\label{eq:nondegeneracyOfG2}
\int_M \dd^dx\,\sqrt{\hat g} \xi^b (G\xi)_b &=& - \int_M\dd^dx\, \sqrt{\hat g}  \left[\xi^b\hat{\nabla}^2{\xi}_b + \frac{2\Lambda}{d-2} \xi^b {\xi}_b+ (1+\alpha) \xi^b \hat{\nabla}_b\hat{\nabla}_a{\xi}^a  \right] \cr
&=& \int_M \dd^dx\,\left[\sqrt{\hat g} (\hat{\nabla}^a {\xi}^b) (\hat{\nabla}_a{\xi}_b) - \frac{2\Lambda}{d-2}\xi^b {\xi}_b+ (1+\alpha) (\hat{\nabla}_b\xi^b) (\hat{\nabla}_a{\xi}^a)\right],
\end{eqnarray}
where in the final step we have used the boundary condition that $\xi^a=0$ on the cavity walls.

Since the spacetime metric has Riemannian signature, the first term on the right-hand side of \eqref{eq:nondegeneracyOfG2} is manifestly positive definite.  In particular,  since $\xi^a=0$ on the cavity walls, this term cannot vanish for any non-zero $\xi^a$.  The final term is similarly positive definite for $\alpha \ge -1$.  Since $\Lambda \le 0$ and $d >2$, the middle term is manifestly non-negative as well.  Thus $G$ is strictly positive-definite (and thus invertible) for all $\alpha \ge -1$, and in this regime there can be no problems with our gauge fixing.

Let us also introduce the operator
\begin{equation}
 K= \mathcal{P}\mathcal{P}^\dagger:\mathfrak{G}\to \mathfrak{G}.
\end{equation}
Note that we have
\begin{equation}
(h, K h)_{\hat{\mathcal{G}}} = (h, \mathcal{P}\mathcal{P}^\dagger h)_{\hat{\mathcal{G}}} =(\mathcal{P}^\dagger h, \mathcal{P}^\dagger h)_{V}.
\end{equation}
Since the inner product on $V$ is positive definite, we see that $K$ annihilates a perturbation $h$ precisely when  $\mathcal{P}^\dagger h=0$; i.e., the kernel of $K$ is the space $W^\perp$.

This suggests that a useful way to lift the degeneracy of $L$ at $\lambda=0$ (see the discussion at the end of section \ref{sec:Wick}) is to define a gauge-symmetry-breaking fluctuation operator $L_{\text{gsb}}$ by
\begin{equation}
\label{eq:gsb}
L_{\text{gsb}} = L + K.
\end{equation}
In particular, $L$ and $L_{\text{gsb}}$ will have identical spectra on $W^\perp$.  Furthermore, while $L$ annihilates $W$, for a complete gauge fixing the operator $L_{\text{gsb}}$ cannot annihilate any perturbation in $W$.  The addition of $K$ would thus completely remove the high degeneracy of $L$ at $\lambda =0$ that results from gauge invariance.  In \eqref{eq:gsb} (as we will shortly explain),  we have chosen the sign of $K$ so that we give positive eigenvalues to the pure gauge modes for $\alpha \ge -2$, though we could in fact have added $K$ with any non-zero coefficient of either sign.

It turns out that the spectra of the operators $G$ and $K$ must coincide for non-zero eigenvalues $\lambda$.  To see this, note that for $G\xi=\lambda\xi$ we have
\begin{equation}
K(\mathcal{P}\xi)=\mathcal{P}\mathcal{P}^\dagger\mathcal{P}\xi=\mathcal{P}G\xi=\lambda(\mathcal{P}\xi),
\end{equation}
while for $Kh= \lambda h$ we have
\begin{equation}
G(\mathcal{P}^\dagger h)=\mathcal{P}^\dagger\mathcal{P}\mathcal{P}^\dagger h=\mathcal{P}^\dagger Kh=\lambda(\mathcal{P}^\dagger h).
\end{equation}
As noted above, pure gauge modes with $Kh=\lambda h$ also satisfy $L_{\text{gsb}} h = \lambda h$.

Furthermore, the sign of an eigenvector of $K$ turns out to always agree with the sign of the norm of the corresponding eigenvector.  In particular, if $h$ is an eigenvector of $K$ with eigenvalue $\lambda$, then we have
\begin{eqnarray}
\label{eq:signlambdanorm}
\lambda (h,h)_{\gma} = (h,\mathcal{P} \mathcal{P}^\dagger h)_{\gma} = (\mathcal{P}^\dagger h, \mathcal{P}^\dagger h)_V \ge 0,
\end{eqnarray}
where the last inequality expresses the fact that we chose a positive-definite inner product on $V$.  We thus see that the sign of any non-zero eigenvalue $\lambda$ is always identical to the sign of the norm $(h,h)_{\gma}$ of the eigenvector $h$. Since  \eqref{eq:nondegeneracyOfG2} showed $G$ to be positive-definite for $\alpha \ge -1$, we see that the non-zero eigenvalues of $K$ are thus also positive for $\alpha \ge -1$.

We can in fact say even more about the spectra of $G$ and $K$.  Let us focus on the action of $G$ on those $\xi^a$ that preserve manifest spherical symmetry; i.e.,  for which $\xi_r(r)= \Xi(r)$ is a function only of $r$, and for which all other components vanish.  Then the condition $G\xi = \lambda \xi$ reduces to
\begin{equation}
\label{eq:Greducedform}
    (2+\alpha)f\Xi''+(2+\alpha)\frac{(d-2)f+2rf'}{r}\Xi'+\frac{-(d-2)(2+\alpha)f+(1+\alpha)r[(d-2)f'+rf'']}{r^2}\Xi=-\lambda\Xi,
\end{equation}
where a prime denotes a derivative with respect to $r$ and we have multiplied by $-1$ to put the left-hand-side in standard form.   If $\Xi_{-1}$ is an eigenfunction  with eigenvalue $\lambda_{-1}$ for $\alpha=-1$, then inserting it into \eqref{eq:Greducedform} as defined by general $\alpha$ shows that $\Xi_{-1}$ remains an eigenfunction of $G$,  but that its eigenvalue becomes
\begin{equation}
\label{eq:Geigenvalues}
\lambda_\alpha=(2+\alpha)\lambda_{-1}-2(1+\alpha)(d-1)/\ell^2.
\end{equation}
In particular, the eigenvalues depend linearly on $\alpha$.  We also see from \eqref{eq:Geigenvalues} that for $\alpha=-2$ {\it all} pure gauge modes have the identical rescaled eigenvalue
\begin{equation}
\label{eq:tildelambda}
\tilde\lambda=\lambda \rp^2=2(d-1)\yp^2 \ge 0.
\end{equation}
And since the eigenvalues $\lambda_{-1}$ must become arbitrarily large at short wavelengths.  The fact that $G$ is positive definite for $\alpha=-1$ then also tells us that the coefficient of $\alpha$ in \eqref{eq:Geigenvalues} is positive, and thus that for any $\alpha < -2$ there will be some negative eigenvalue $\lambda_{\alpha}$.  In particular, on a given background the operator $G$ will fail to be invertible at a sequence of $\alpha$-values that converge to $\alpha=-2$ from below.  In contrast, we see that $G$ is positive-definite (and thus invertible) for all $\alpha > -2$ (and also for $\alpha=-2$ when $\Lambda < 0$).  Thus the spectrum of $K$ is also positive-definite for $\alpha > -2$) (and for $\alpha=-2$ when $\Lambda < 0$).

It is worth remarking that the operator $G$ fails to be elliptic precisely in the regime $\alpha<-2$ where it fails to be positive definite.  This should not be a surprise, as the operator is in fact hyperbolic for $\alpha<-2$, so that its spectrum cannot have a definite sign.  In much the same way, one can show that the gradient flow defined by the action and some DeWitt$_\alpha$ metric is parabolic when linearized about a saddle for $\alpha > -2$, but that it fails to become parabolic for $\alpha < -2$.

\subsection{Boundary Conditions}\label{sec:boundaryCondition}

The spherical symmetry of our background spacetime makes it is natural to expand any perturbation $h_{ab}$ in terms of spherical harmonics.  Furthermore,  since stability of a saddle is determined by the lowest eigenvalue of $L_\alpha$, the fact that angular momentum contributes positively to \eqref{eq:Lalpha} implies that we need only analyze the modes with lowest angular momentum.  In particular, from \eqref{eq:Lalpha} we see that any $\alpha$-dependence of angular momentum contributions will arise only from the trace part of the operator $L$, and that the trace part will maintain the same (positive) sign as for $\alpha = -1$ so long as $\alpha < -1/(d-1)$. Since $d  >2$ we have $-2/d < -1/(d-1)$. We thus find that the contribution of angular momentum is always positive in the regime $\alpha < - 2/d$ where the DeWitt$_\alpha$ metric has indefinite signature.
As a result, we focus on spherically symmetric perturbations below\footnote{In principle, the lowest vector and tensor harmonics should be checked separately.  However, we leave this for future study.}.

As in \cite{Marolf:2022ntb}, for spherically symmetric perturbations imposing the $\tau$ component of our gauge condition (\ref{eq:newdedonder}) requires
\begin{equation}
\label{eq:fixtauparts}
h_{r\tau}=h_{\tau r}=0.
\end{equation}
We will impose \eqref{eq:fixtauparts} below for all $\alpha$, which means that (in effect) we henceforth take the space $\mathfrak{G}$ to include only perturbations satisfying \eqref{eq:fixtauparts}.

A general such perturbation can be written in the form
\begin{equation}\label{eq:perturbation}
    \delta\dd s^2=a(r)f(r)\dd\tau^2+\frac{b(r)}{f(r)}\dd r^2+c(r) r^2\dd \Omega_{d-2}^2.
\end{equation}
In the canonical ensemble,  the perturbation is required to preserve the induced metric at the cavity wall.  Since the wall is located at $r=r_0$, we thus find\begin{equation}\label{eq:dirichletBC}
    a(r_0)=c(r_0)=0.
\end{equation}

However, as mentioned in section \ref{sec:gaugeCondition}, for finite $r_0$ we must impose an additional boundary condition requiring \eqref{eq:Pdagonh} to vanish at $r=r_0$. To write this additional condition in a more explicit form,  let us first decompose any perturbation \eqref{eq:perturbation} into a pure-trace part (given by $p$ times the background metic)   and a trace-free part (given by \eqref{eq:perturbation} with $\hat a$, $\hat b$, $\hat c$ satisfying $\hat a + \hat b +(d-2)\hat c=0$, so that we find
\begin{equation}
    a=\frac{p}{d}-\hat{b}-(d-2)\hat{c},\quad b=\hat{b}+\frac{p}{d},\quad c=\hat{c}+\frac{p}{d}.
\end{equation}
After imposing $h_{r\tau}=0$, the gauge condition (\ref{eq:newdedonder}) becomes equivalent to the first-order ordinary differential equation
\begin{equation}
\label{eq:gcode}
    -(d-2)(2f-rf')\hat{c}+2rf\hat{b}'+2[(d-2)f+rf']\hat{b}+\frac{(2+d\alpha)rf}{d}p'=0,
\end{equation}
where a prime denotes a derivative with respect to $r$. Solving for $\hat c$ then yields
\begin{equation}
    \hat{c}=\frac{2d[(d-2)f+rf']\hat{b}+2drf\hat{b}'+(2+\alpha d)rfp'}{d(d-2)(2f-rf')}.
\end{equation}
Setting $r=r_0$ and using (\ref{eq:dirichletBC}), we find that \eqref{eq:gcode} becomes
\begin{equation}\label{eq:BCgauge}
    p'(\ro)+\frac{d f'(\ro)p(\ro)}{(2+d\alpha)f(\ro)}+\frac{(2d^2-4d) p(\ro)}{(2+d\alpha)\ro}+\frac{2d\hat{b}'(\ro)}{2+d\alpha}=0.
\end{equation}

We will also need to impose boundary conditions at the horizon $r=\rp$. Here we simply require the perturbations to be regular.

In performing numerical calculations, it is often convenient to work with a coordinate that ranges only over a closed finite interval.  We thus define the coordinate $y\in [0,1]$ by setting
\begin{equation}\label{eq:ydef}
    y=\sqrt{\frac{1-\rp/r}{1-\rp/\ro}}.
\end{equation}
In terms of the new coordinate $y$, the boundary conditions at the cavity wall $y=1$ take the form
\begin{equation}\label{eq:bc1}
\begin{split}
    &a(1)=c(1)=0,\\
    d\alpha a'(1)+(\alpha+2)db'(1)+d\alpha(d-2) &c'(1)+2\left(\frac{\ro}{\rp}-1\right)\left[\frac{d\ro f'(\ro)}{f(\ro)}+2d(d-2)\right]b(1)=0.
\end{split}
\end{equation}
Similarly,
regularity at the horizon $y=0$ is equivalent to the conditions
\begin{equation}\label{eq:bc2}
   \quad a(0)=b(0),\quad\text{and}\quad   a'(0)=b'(0)=c'(0)=0.
\end{equation}

\subsection{Numerical Scheme}\label{sec:scheme}

With the correct gauge condition and the original quadratic action in hand, we are now ready to reduce the degeneracy of $L$ at $\lambda=0$ by adding a gauge-symmetry breaking term.  We may do so by defining a new action $\check{S}^{[2]}[h]$ by adding $(h,K \tilde{h})_{\gma}$ to the quadratic action $S^{[2]}[h]$; i.e.,
\begin{equation}\label{eq:newaction}
    \check{S}^{[2]}[h] :=S^{[2]}[h]+(h,K{h})_{\gma}.
\end{equation}

The numerical scheme we will use will be identical to that used in \cite{Marolf:2022ntb}, and is based on discretizing the action \eqref{eq:newaction} and then using the discrete analogue of  \eqref{eq:Ldef}  to define a discretized version of $L_{\text{gsb}}$.  Before briefly summarizing this scheme we should mention that, if we had used the traditional operator approach \cite{Gibbons:1978ji,Gross:1982cv,Allen:1984bp,Monteiro:2009ke,Monteiro:2009tc,Dias:2009iu,Prestidge:1999uq,Dias:2010eu,Headrick:2006ti}, the operator we discretize on a numerical grid would not be $\hat{\Delta}_L$ because $S^{[2]}\neq (h,\hat{\Delta}_L h)_{\gma}$ if $\alpha\neq -1$. Instead, we have
\begin{equation}
    \check{S}^{[2]}=(h,L_{\alpha}h)_{\gma}+(h,K{h})_{\gma}.
\end{equation}
As a result, the eigenvalue problem to be solved is
\begin{equation}
(L_{\text{gsb}, \alpha}h)_{ab} : =     \left[ (L_\alpha +K) h\right]_{ab}=\lambda h_{ab}.
\end{equation}

Working with the coordinate $y\in[0,1]$, we use spectral collocation methods on Gauss-Lobatto collocation points to define $N+1$ grid points:
\begin{equation}
    y_i=\frac{1}{2}\left[1+\cos\frac{i\pi}{N}\right],\quad i=0,\cdots,N.
\end{equation}
As a result, $y_0=1$ corresponds to the location of the cavity wall, while $y_{N}=0$ corresponds to the location of the horizon. We also define $\mathbb{f}_i\equiv f(y_{i-1})$ as the discretized version of any function $f$. Numerical approximation of differentiation and integration of functions can be performed by matrix multiplication on the vector $\vec{f}\equiv(\mathbb{f}_1,\cdots,\mathbb{f}_{N+1})$, where the relevant reflection matrices can be found in standard textbooks or review articles \cite{Dias:2015nua, Canuto2010SpectralMF, boyd2001chebyshev}.

Since there are $3$ functions ($a,b$ and $c$) in our perturbation, the discretized field variable $\tilde{\mathbb{Q}}$ has $3(N+1)$ components $\tilde{\mathbb{Q}}\equiv(\mathbb{a}_1,\cdots,\mathbb{a}_{N+1},\mathbb{b}_1,\cdots,\mathbb{b}_{N+1},\mathbb{c}_1,\cdots,\mathbb{c}_{N+1})$ (though due to the boundary conditions we will impose below, these components will not be completely independent). The modified quadratic action can then be written in the form
\begin{equation}\label{eq:discretizedAction}
    \check{S}\approx \check{\mathbb{S}}^{[2]}=\sum_{I,J=1}^{3(N+1)}\tilde{\mathbb{Q}}^{I}\tilde{M}_{IJ}\tilde{\mathbb{Q}}^{J}.
\end{equation}

The next step is to discretize the boundary conditions and derive the associated constraints on $\tilde{\mathbb{Q}}$. The discretized version of our boundary conditions (\ref{eq:bc1}) and (\ref{eq:bc2}) are
\begin{equation}
    \begin{split}
        \mathbb{a}_1=\mathbb{c}_1=0,\quad \mathbb{a}_{N+1}=\mathbb{b}_{N+1},\quad &\mathbb{D}_{N+1}\cdot\vec{\mathbb{a}}=\mathbb{D}_{N+1}\cdot\vec{\mathbb{b}}=\mathbb{D}_{N+1}\cdot\vec{\mathbb{c}}=0\\
    d\alpha\mathbb{D}_{1}\cdot\vec{\mathbb{a}}+(\alpha+2)d\mathbb{D}_{1}\cdot\vec{\mathbb{b}}+d\alpha(d-2)\mathbb{D}_{1}&\cdot\vec{\mathbb{c}}+2\left(\frac{\ro}{\rp}-1\right)\left[\frac{d\ro f'(\ro)}{f\ro)}+2d(d-2)\right]\mathbb{b}_1=0.
    \end{split}
\end{equation}
Here $\mathbb{D}$ is the differentiation matrix and the subscript indicates which row we are using. We now solve the above $7$ equations for $\{\mba_1,\mba_N,\mba_{N+1},\mbb_1,\mbb_{N+1},\mbc_1,\mbc_{N+1}\}$ and insert the solutions into equation (\ref{eq:discretizedAction}) to find a discretized action (with a gauge-symmetry-breaking term) in terms of unconstrained variables $\mathbb{Q}\equiv(\mathbb{a}_2,\cdots,\mathbb{a}_{N-1},\mathbb{b}_2,\cdots,\mathbb{b}_{N},\mathbb{c}_2,\cdots,\mathbb{c}_{N})$:
\begin{equation}\label{eq:dicretizedAction}
    \check{\mathbb{S}}=\sum_{I,J=1}^{3N-4} \mathbb{Q}^I\check{S}_{,IJ}\mathbb{Q}^J.
\end{equation}
Following the same procedure, we can discretize the norm of $h_{ab}$ to construct a discretized version of $\gma$:
\begin{equation}\label{eq:norm}
    \|h\|^2\approx\sum_{I,J=1}^{3N-4} \mathbb{Q}^I{\mathbb{G}}_{IJ}\mathbb{Q}^J.
\end{equation}

The numerical problem we wish to solve is then the generalized eigenvalue problem
\begin{equation}
\label{eq:discreteEqs}
    \mbl \cdot \mathbb{Q}_\lambda=\lambda\hat{\mathbb{G}}\cdot\mathbb{Q}_{\lambda},\quad \text{where}\quad \mbl{}^I_{\,J}=\sum_K \delta^{IK}\check{\mathbb{S}}_{,IJ},\quad \text{and}\quad \hat{\mathbb{G}}^{I}_{\,J}=\sum_K \delta^{IK}\mathbb{G}_{IJ},
\end{equation}
and the notation $\cdot$ indicates the natural action of a matrix as a linear map on the space of vectors.
Note that the definition of $\mathbb{L}$ in \eqref{eq:discreteEqs} is just the discretized analogue of \eqref{eq:Ldef}.  In particular, it guarantees $\mathbb{L}$ to be self-adjoint with respect to the discretized metric $\hat{\mathbb{G}}$.

After numerically obtaining an eigenmode, we will need to decide whether it is a physical mode or a pure gauge mode. This can be checked by computing the corresponding gauge-symmetry breaking term $(h,K h)_{\gma}$ using a similar discretized form analogous to equations (\ref{eq:discretizedAction}) and (\ref{eq:norm}). If the result is zero, then this mode is a physical mode.  Otherwise, the mode is a pure gauge. As a consistency check, we can also compute the corresponding unmodified action $S^{[2]}$. If it is zero, then the mode is a pure gauge mode, otherwise, it is a physical mode.

\section{Results and Interpretation}
\label{sec:NumericalResults}
We now present our numerical results regarding mode stability for different choices of the parameter $\alpha$ in the DeWitt metric.  We focus on results for perturbations that preserve manifest  time-translation invariance and spherical symmetry. The results are expressed in terms of the dimensionless quantities $\yo=\ro/\rp,\ \yp=\rp/\ell$ and $\tilde{\lambda}\equiv \lambda\rp^2$.  As described above,  if $\alpha>-2/d$, the DeWitt metric will be positive definite so that the rule-of-thumb states that no Wick rotation should be performed. However,  since the conformal factor problem makes the action unbounded below\footnote{The problem is local and so occurs for any choice of boundary conditions.}, there must be modes for which the action is negative.  Thus, using the rule-of-thumb, the path integral can have no stable saddles (regardless of whether any solution is thermodynamically stable).  This is a clear failure of the rule-of-thumb for $\alpha \ge -2/d$.  We thus consider only $\alpha < - 2/d$ below.    We focus on $d=4$, but we expect similar results in higher dimensions (or with the addition of bulk matter for $d \ge 3$).  The case $\alpha > -2$ (where $G$ is positive definite) is described in section \ref{sec:Gposdef}, while the case $\alpha <-2$ (where $G$ fails to be positive definite) is discussed in section \ref{sec:Gnotposdef}.

\subsection{The case $-2/d>\alpha>-2$ for $d=4$}
\label{sec:Gposdef}

Since the stability of saddles is determined by the existence of negative modes, we first focus on the lowest eigenmode of our fluctuation operator.  Section \ref{sec:lowest} reports results for $-2/d > \alpha > -2$ with $d=4$. The results are described in terms of the dimensionless eigenvalue $\tilde \lambda$ defined in \eqref{eq:tildelambda}.  The excited spectrum is then discussed in section \ref{sec:bubble}.  In particular, we revisit the transitions between real and complex eigenvalues described in \cite{Marolf:2022ntb}.  Diagonalizability of the operator $L$ turns out to fail at such transitions, so that the rule-of-thumb of \cite{Marolf:2022ntb} fails as well.  However, section \ref{sec:contours} then argues these breakdowns to be harmless as the contours defined on either side of such transitions by the rule-of-thumb are continuous deformations of each other (so that, since our integrand is purely Gaussian and thus cannot have poles, Cauchy's theorem states that they define the same path integral).

\subsubsection{The lowest eigenvalue and its sign}
\label{sec:lowest}

The value $\alpha=-1$ was studied in \cite{Marolf:2022ntb}, which found that the lowest value of $\tilde{\lambda}$ changed sign precisely on the curve determined by  (\ref{eq:ypstar}), with positive values of $\tilde{\lambda}$ occurring at smaller values of $y_+$ (which correspond to larger values of $r_+$).  In addition, \cite{Marolf:2022ntb} found all modes with smallest $\Re \tilde{\lambda}$ to be real\footnote{While we do not have an analytic proof of this statement, it is consistent with the fact that \cite{Headrick:2006ti} found only exponential growth (and not oscillations) in the corresponding Ricci flow simulation.}, and to have positive norm.  For fixed $\yp$, the eigenvalue for the lowest mode was also found to decrease as $\yo$ increases.

\begin{figure}[h!]
     \centering
     \begin{subfigure}{0.32\textwidth}
         \includegraphics[width=\linewidth]{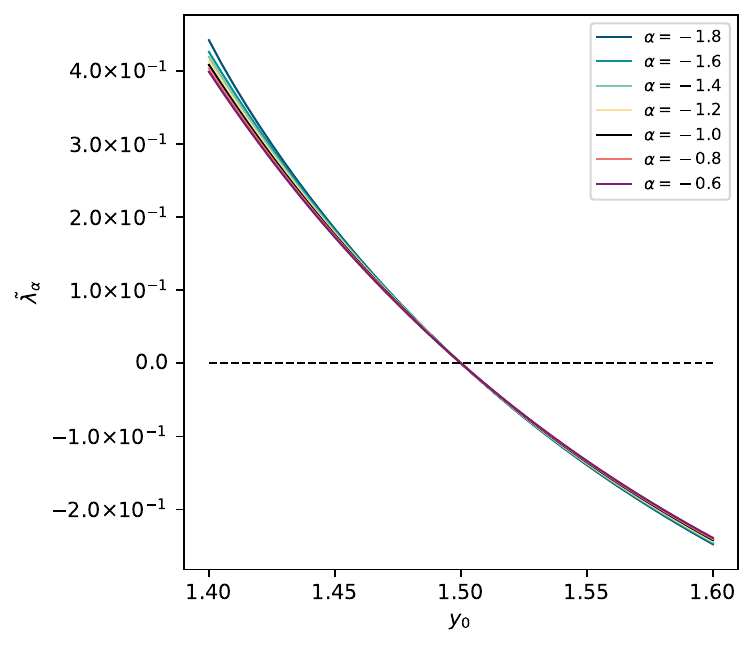}
         \caption{$\yp=0$}
         \label{fig: lowest_modes_a}
     \end{subfigure}
     \begin{subfigure}{0.32\textwidth}
         \includegraphics[width=\linewidth]{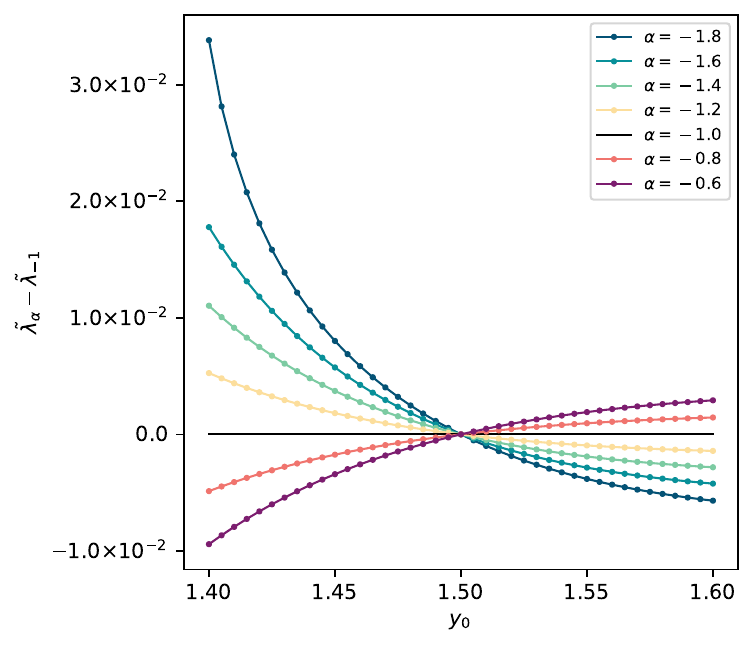}
         \caption{$\yp=0$}
         \label{fig: lowest_modes_b}
     \end{subfigure}
     \begin{subfigure}{0.32\textwidth}
         \includegraphics[width=\linewidth]{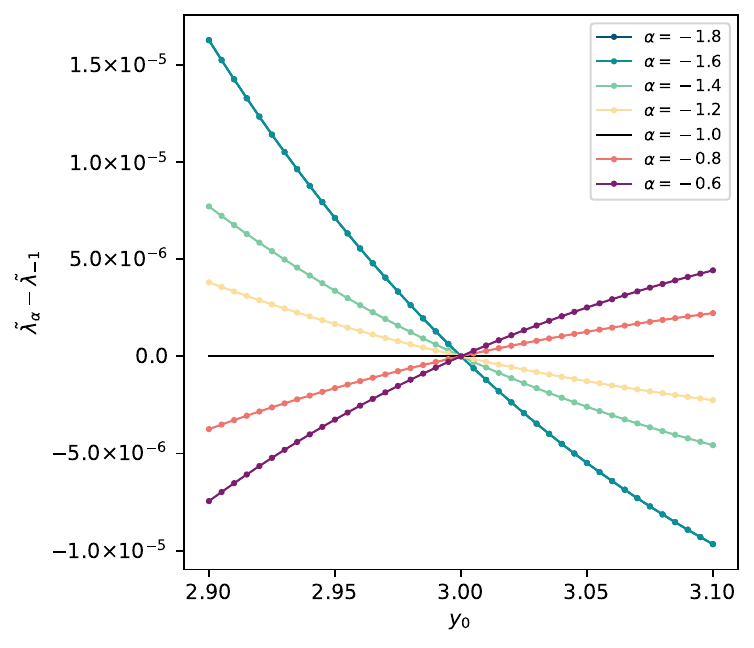}
         \caption{$\yp=\sqrt{3/11}$}
         \label{fig: lowest_modes_c}
     \end{subfigure}
        \caption{The lowest dimensionless eigenvalue $\tilde \lambda$ is plotted in panel (a) for $y_+=0$ as a function of $\alpha$.  The dashed line indicates $\tilde\lambda=0$, and the data agrees well with the known change in thermodynamic stability at $\yo=1.5$.  Panel (b) shows the same data, but with the vertical axis now taken to be the difference between the dimensionless eigenvalue at each given $\alpha$ and the value at $\alpha=-1$.  Panel (b) also uses an enlarged vertical scale.   Panel (c) shows the analogue of (b) for $\yp = \sqrt{3/11}$  (with an even more enlarged scale).  For this value of $\yp$ the transition in thermodynamic stability occurs at $\yo=3$, which agrees well with the data shown.    In both cases we take $d=4$ so that we focus on the range of $\alpha$ between $-2/d=-1/2$ and $-2$.}
        \label{fig:lowest_modes}
\end{figure}
These features turn out to hold for all $\alpha$ in the present range $-2/d>\alpha>-2$. This is illustrated in Figure \ref{fig: lowest_modes_a} for $\yp=0$.  Since the lowest eigenvalue changes only very slowly with $\alpha$, the same data is also plotted again in figure \ref{fig: lowest_modes_b}  using an enlarged scale and taking the vertical axis to be the difference between the dimensionless eigenvalue $\tilde \lambda_\alpha$ at the given $\alpha$ and the dimensionless eigenvalue $\tilde \lambda_{-1}$ associated with the value $\alpha=-1$.  We also plot this quantity in figure \ref{fig: lowest_modes_c} for the anti-de Sitter case $\yp=\sqrt{3/11}$, where \eqref{eq:ypstar} places the transition at precisely $\yo=3$.  We find similar results for other values of $\yp>0$. In all cases, it appears that -- as expected from the discussion at the end of section \ref{sec:stabcomp} -- the curves describing eigenvalues at different $\alpha$ all appear to cross $\tilde \lambda =0$ at at exactly the value of $\yo$ where the specific heat changes sign. We also find the lowest modes at all $\alpha \in (-2, -2/d)$ to have positive norms, so that they should not be Wick-rotated.   Our $d=4$ data for the lowest mode thus supports the idea that the rule-of-thumb succeeds for $-2/d > \alpha > -2$.

\begin{figure}[t!]
     \centering
     \begin{subfigure}{0.49\textwidth}
         \includegraphics[width=\linewidth]{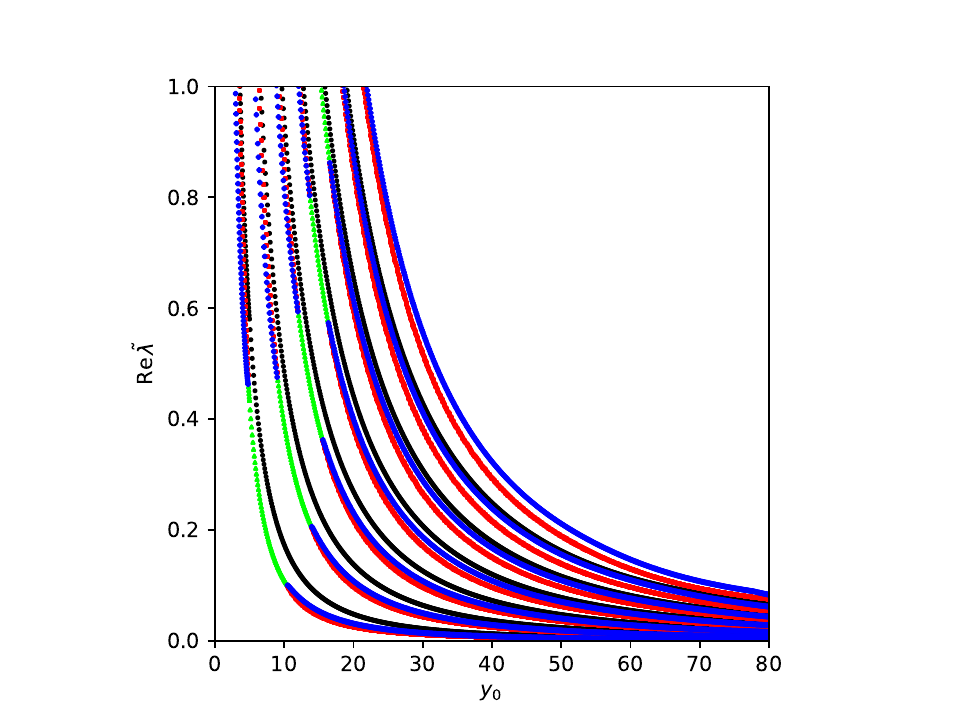}
         \caption{$\alpha=-0.95$}
         \label{fig: alpha0.95mergeRe}
     \end{subfigure}
     \begin{subfigure}{0.49\textwidth}
         \includegraphics[width=\linewidth]{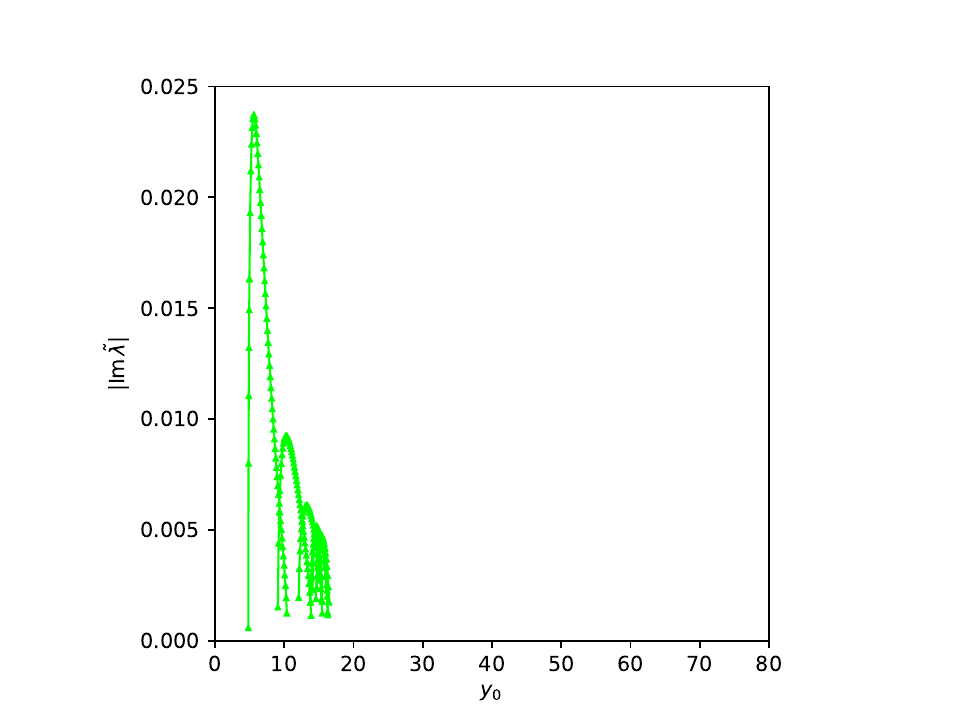}
         \caption{$\alpha=-0.95$}
     \end{subfigure}
     \begin{subfigure}{0.49\textwidth}
    \includegraphics[width=\linewidth]{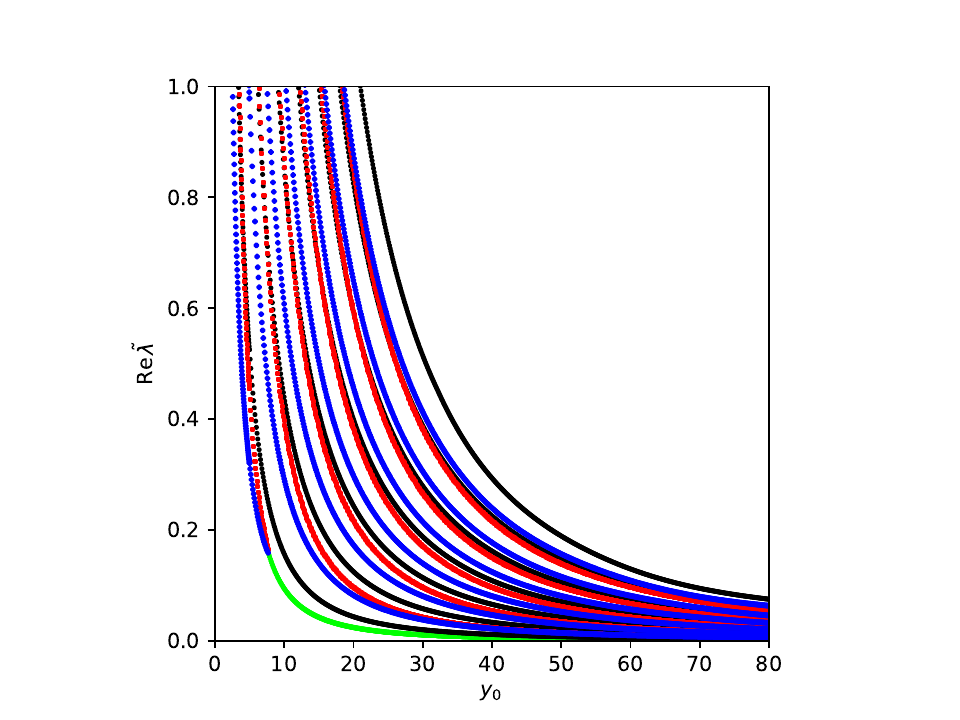}
         \caption{$\alpha=-1.05$}
     \end{subfigure}
     \begin{subfigure}{0.49\textwidth}
    \includegraphics[width=\linewidth]{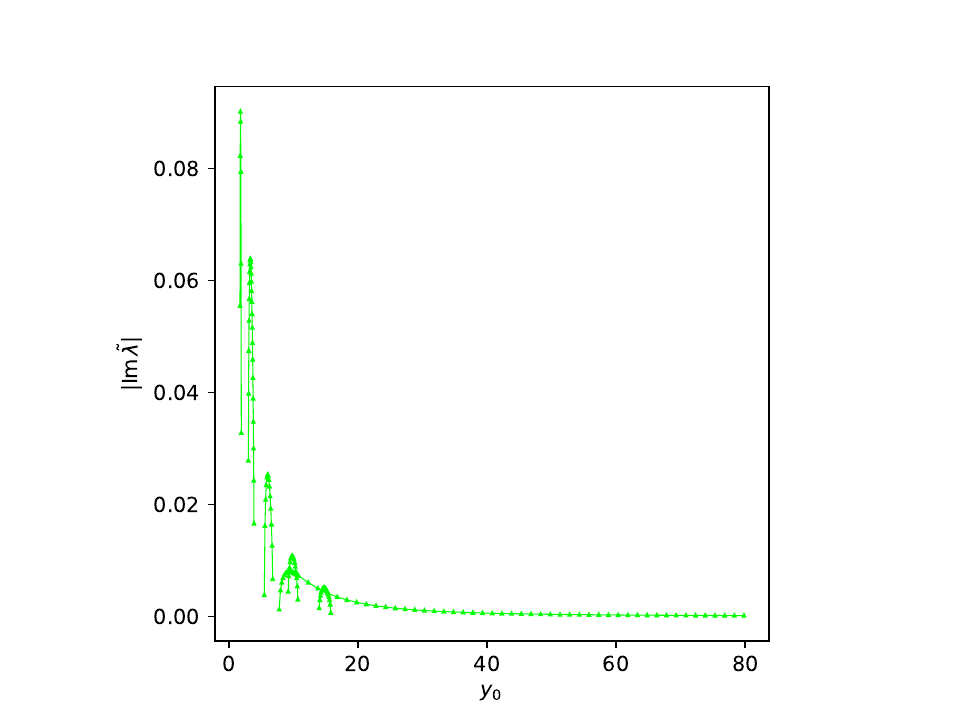}
         \caption{$\alpha=-1.05$}
     \end{subfigure}
        \caption{The real part (left column) and the magnitude of the imaginary part (right column) of the first $20$ excited-mode eigenvalues are shown as functions of $\yo$ in the case $\Lambda=0$. The eigenvalues of pure gauge modes (black) are always real and their eigenvectors have positive norm.  Positive-norm (negative-norm) physical modes with real eigenvalues are shown in red (blue).  Physical modes with complex eigenvalues are shown in green, with each green data point representing a pair of complex-conjugate  modes with complex-conjugate eigenvalues. The only complex modes from panel (d) that ccan be seen in figure (c) are in the family that runs to $\yo=80$ and beyond. The others have $\rm{Re}\,\tilde \lambda >1$ and so do not appear in figure (c).}
        \label{fig:firstExcitedModes}
\end{figure}

\subsubsection{The excited spectrum: failure of diagonalizability at bubble walls}
\label{sec:bubble}

Recall that the rule-of-thumb is well-defined only when $L$ is diagonalizable, or equivalently when $L_{\text{gsb}}$ can be diagonalized on the space of physical modes.  This property is clearly a statement about the complete spectrum of the operators and not just their lowest eigenvalues.  It is thus important to investigate the excited spectrum in addition to the lowest eigenvalue already described in section \ref{sec:lowest}.

Now, as described in \cite{Marolf:2022ntb}, for the case $\alpha=-1$ the excited spectrum displays regions of parameter space in which certain eigenvalues are complex.  These regions were called `complex bubbles,' and arise when a real-eigenvalue positive-norm physical mode becomes degenerate with a real-eigenvalue negative-norm physical mode.  Recall that such transitions are allowed due to the indefinite signature of the metric with respect to which $L$ and $L_{gsb}$ are self-adjoint.  Such transitions were briefly discussed in section \ref{sec:Wick}, where it was noted that they generically render the operator non-diagonalizable.

Such behavior  would imply a break-down of the rule-of-thumb.  Indeed, it was noted in section \ref{sec:Wick} that, if we study the two eigenmodes away from the transition and then take the limit as the transition is approached from either side, the limiting eigenmodes will coincide.  But when the transition is approached from the real-eigenvalue side, one of the eigenmodes is positive-norm and the other is negative-norm.  Thus the rule of thumb rotates one of the modes while leaving the other invariant.  Thus there can be no well-defined limit of such contours at the transition where the two modes coincide.  This again suggests that our operator will simply fail to be diagonalizable at the transition itself (i.e., on the walls of the complex bubbles).

We will verify this behavior in detail below.  However, we will also show such failures to be harmless in the following sense.  Although the rule-of-thumb breaks down at the transition, and although the limit of rule-of-thumb contours at the transition is not well-defined, we nevertheless find that contours on opposite sides of the wall can be continuously deformed to each other.    Since our integral is purely Gaussian (and thus cannot have poles), Cauchy's theorem then states that the two classes of contours define the same path integral.

We begin with results for the excited modes and complex bubbles.
Figure \ref{fig:firstExcitedModes} depicts our $\yp=0$ data for the first 20 excited modes as a function of $\yo$ in the two cases $\alpha=-0.95$ and $\alpha=-1.05$.  We see that all modes have $\Re \tilde\lambda>0$, thus all the excited modes are stable. For certain ranges of parameters we also find that some modes are complex, having eigenvalues with a nonvanishing imaginary part.  These results are similar to the $\alpha=-1$ results reported in \cite{Marolf:2022ntb}.  We find similar results for $\yp>0$ and other values of $\alpha$.

\begin{figure}[!h]
\centering
\includegraphics[width=\linewidth]{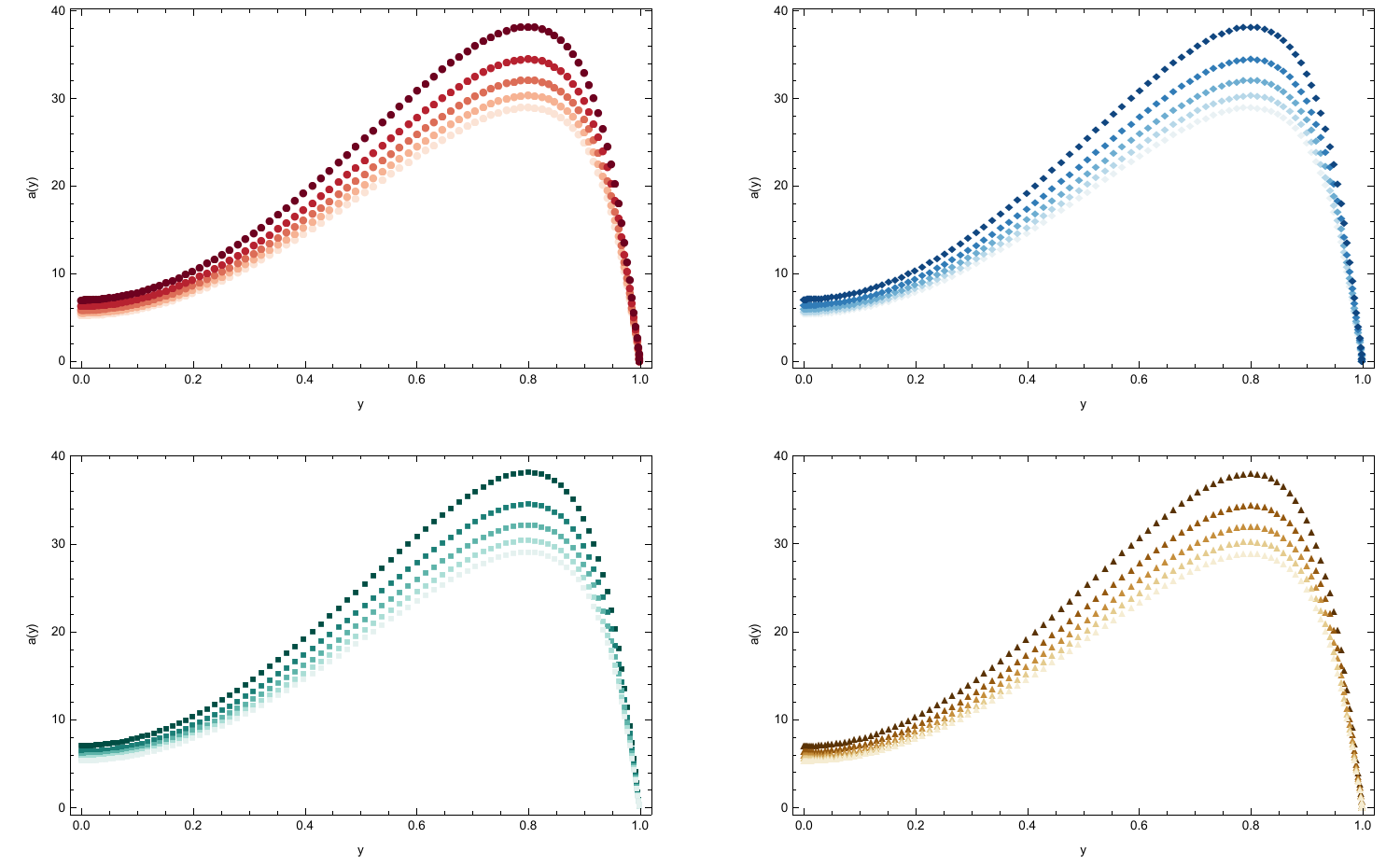}
\caption{The functions $a(y)$ for merging modes at values of $\yo$ close to a bubble wall. {\bf Top:} The left/right panel shows a real physical mode with positive/negative norm. {\bf Bottom:} The left/right panel shows the real/imaginary part of the complex mode. Here $\alpha=-1$ and $\yp=0$. In the upper panels, from light to dark, the values of $\yo$ are $6.028489,\ 6.028490,\ 6.028491,\ 6.028492,\ 6.028493$. In the lower panel, from dark to light, the values of $\yo$ are $6.028495,\ 6.028496,\ 6.028497,\ 6.028498,\ 6.028499$. Darker colors are closer to the wall (at $\yo \sim 6.028494$) and show larger amplitudes due to the eigenvectors becoming null. Our parameter values are close to being symmetric about the edge of the bubble so that the eigenfunctions on opposite sides agree to a good precision.  }
\label{fig:modes_bubble}
\end{figure}

We can verify numerically that the relevant two eigenmodes tend to the same limiting mode function at the bubble wall.  In fact, we can do so in two different ways.  The first is shown in figure \ref{fig:modes_bubble}, which simply plots the functions $a(y)$ for (on the real side of the wall) the relevant positive- and negative-norm eigenvector at parameters close to a wall, as well as (on the complex side of the wall) the real and imaginary parts of the relevant complex eigenvector at nearby parameters.  The results are visually indistinguishable, supporting the claim that all 4 eigenvectors approach the same perturbation at the wall.  This in fact was seen previously in \cite{Marolf:2022ntb}, though the full implications were not understood.

\vspace{.5cm}
\begin{figure}[!h]
    \centering
    \begin{subfigure}{0.45\textwidth}
        \includegraphics[width=\linewidth]{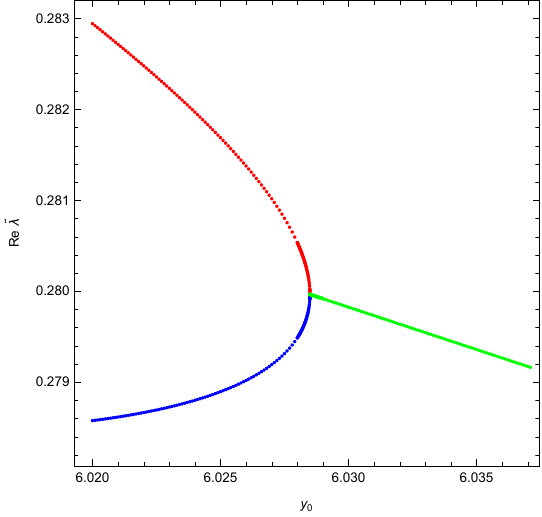}
        \caption{}
        \label{fig:evals_bubble}
    \end{subfigure}
    \hspace{1.5cm}
    \begin{subfigure}{0.45\textwidth}
        \includegraphics[width=\linewidth]{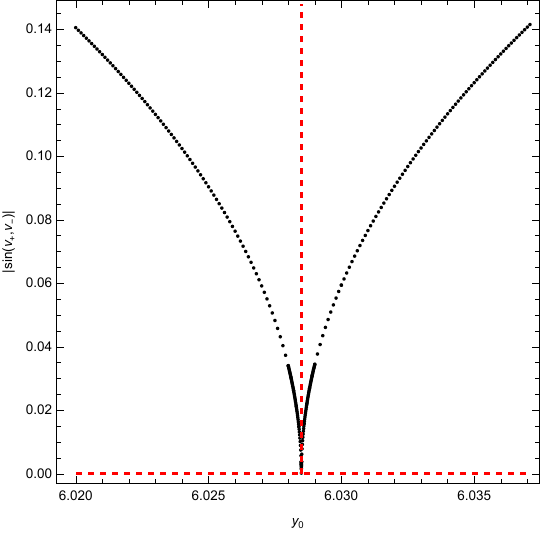}
        \caption{}
        \label{fig:angle_bubble}
    \end{subfigure}
    \caption{The left panel shows the real part of a pair of eigenvalues that merge at a bubble wall.  The red/blue data points are real physical eigenvalues with positive/negative eigenvectors. Each green dot represents a complex-conjugate pair. The right panel shows the absolute value of the sine of the angle (defined by the Cartesian metric \eqref{eq:CartesianMetric}) between the relevant two modes (the two real eigenvectors for $y_0<y_0^*$ and the real and imaginary parts of the complex eigenvector for $y_0>y_0^*$). The red dashed horizontal line indicates $|\sin (v_1,v_2)|=0$. The vertical red dashed line indicates the bubble wall. Here $\alpha=-1$ and $\yo=0$.}
    \label{fig:angle_bubble_m1}
\end{figure}

We can also perform a quantitative check that e.g. the positive- and negative-norm eigenvectors (on the real side of the wall) coincide at the wall.  To do so, we define an angle between the two modes by using the fixed Cartesian inner product
\begin{equation}
	\label{eq:CartesianMetric}
	(\mathbb{Q},{\mathbb{Q}'})_{\mathrm{C}}=\mathbb{Q}^*\cdot {\mathbb{Q}'}=\mathbb{a}^*_2{\mathbb{a}'}_2+\cdots+\mathbb{a}^*_{N-1}{\mathbb{a}'}_{N-1}+\mathbb{b}^*_2{\mathbb{b}'}_2+\cdots+\mathbb{b}^*_{N}{\mathbb{b}'}_{N}+\cdots+\mathbb{c}^*_{N}{\mathbb{c}'}_{N},
\end{equation}
where $^*$ denotes complex conjugation, $\mathbb{Q}$ was defined as above equation \eqref{eq:dicretizedAction}, and ${\mathbb{Q}'}$ is another vector defined analogously.
  As shown in figure \ref{fig:angle_bubble_m1} (right), we find numerically that this angle vanishes at the transition.  For reference, the left panel is a zoomed-in version of figure of figure 3 in \cite{Marolf:2022ntb} showing the particular transition plotted on the right.

As described in section \ref{sec:scheme}, our discretized problem involves a space of perturbations of finite dimension $(3N-4)$.  And away from the wall we can identify $(3N-4)$ eigenvectors.  The fact that two eigenvectors coincide at the wall thus suggests that there can be only $(3N-5)$ eigenvectors at the wall, so that our operator fails to be diagonalizable.

However, we should verify that a new eigenvector does not spontaneously arise when the wall is reached.  As described in section \ref{sec:Wick}, near the transition we expect that we can describe the relevant eigenvectors as effectively evolving in a fixed two-dimensional space.  (We will verify this numerically below in figure \ref{fig:comp_div_bubble}.)  As we approach the bubble wall, the two eigenvectors approach a common real null vector as described in section \ref{sec:Wick}.  Let us call this vector $v_1$, so that
\begin{equation}
\label{eq:v1eq}
L v_1  = \lambda v_1
\end{equation}
at the transition.  As also discussed in section \ref{sec:Wick}, it is of interest to discuss the conjugate null vector $v_2$ in this two-dimensional space (satisfying $(v_2,v_1)_{\gma}=1$).  On general grounds this $v_2$ should satisfy \eqref{eq:gamma}, which for the convenience of the reader we reproduce here:
\begin{equation}
\label{eq:gamma2}
L v_2  = \lambda v_2 + \gamma v_1 .
\end{equation}
Recall that the $\lambda$ in \eqref{eq:gamma2} takes the same value as in \eqref{eq:v1eq}.  When $\gamma \neq 0$, the operator is not diagonalizable.

We can verify numerically that this is the case by considering parameters $p$ close to the wall and constructing the real vectors $v_1, v_2$ that are exactly null as determined by each separate $p$.  Since we are not exactly at the wall, neither of these will be precisely an eigenvector.  But $v_1$ should be close to being so.  We then define the four quantities
\begin{eqnarray}
\label{eq:betagamma}
\lambda_1 = (v_2, L v_1)_{\gma}, \ \ \ \lambda_2 = (v_1, L v_2)_{\gma} \cr
\beta = (v_1, L v_1)_{\gma}, \ \ \ \gamma = (v_2, L v_2)_{\gma}.
\end{eqnarray}

Self-adjointness of $L$ requires $\lambda_1 = \lambda_2^*$ and also requires $\beta, \gamma$ to be real, but there is no general relation between $\beta$ and $\gamma$.  Indeed, requiring $(v_1,v_2)_{\hat{\mathcal{G}}_\alpha}=1$ does not fix the normalization of either $v_{1}$ or $v_{2}$ separately. For example, if we are on the left-hand side of the bubble in figure \ref{fig:evals_bubble} and we have the $\hat{\mathcal{G}}_{-1}$-normalized eigenvectors $v_\pm$ of positive and negative norm, then in general we allow $v_1=\mathcal{N}(v_++v_-)/\sqrt{2}$ and $v_2=\mathcal{N}^{-1}(v_+-v_-)/\sqrt{2}$.    Here $\mathcal{N}$ is an arbitrary $p$-dependent normalization constant, though it must vanish at transition if $v_1$ is to remain finite in the limit where $v_\pm$ become null.  It is thus useful to choose $\mathcal{N}$ at each $p$ so that $v_1$ has norm $1$ with respect to the fixed Euclidean-signature metric  \eqref{eq:CartesianMetric}.

\begin{figure}[b!]
    \centering
    \begin{subfigure}{0.32\textwidth}
    	\includegraphics[width=\linewidth]{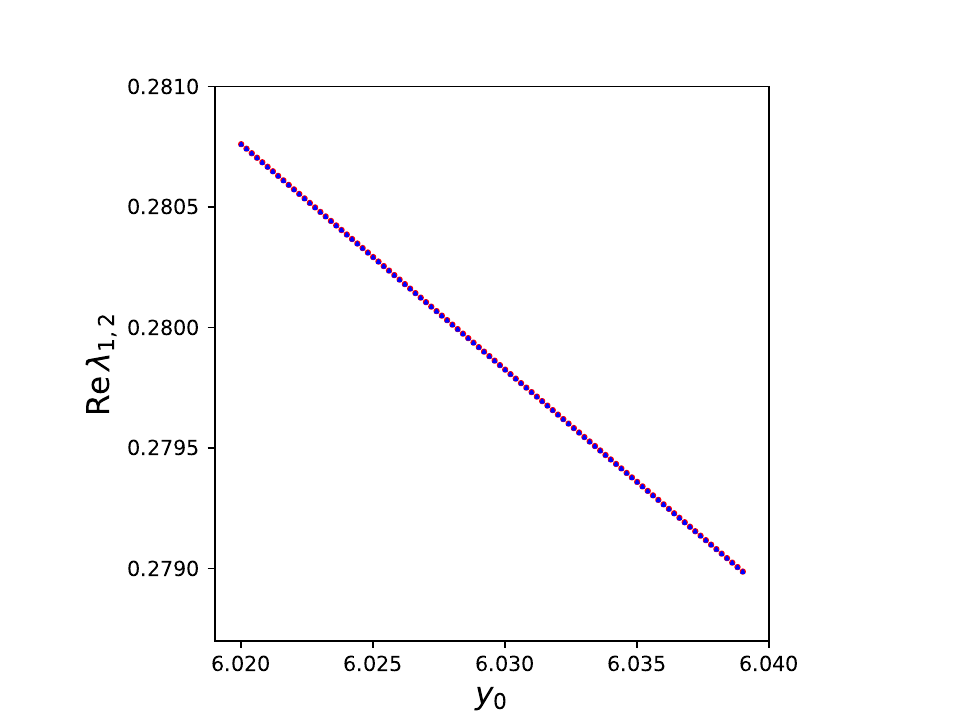}
    	\caption{}
    	\label{fig:lambda_bubble}
    \end{subfigure}
    \begin{subfigure}{0.32\textwidth}
    	\includegraphics[width=\linewidth]{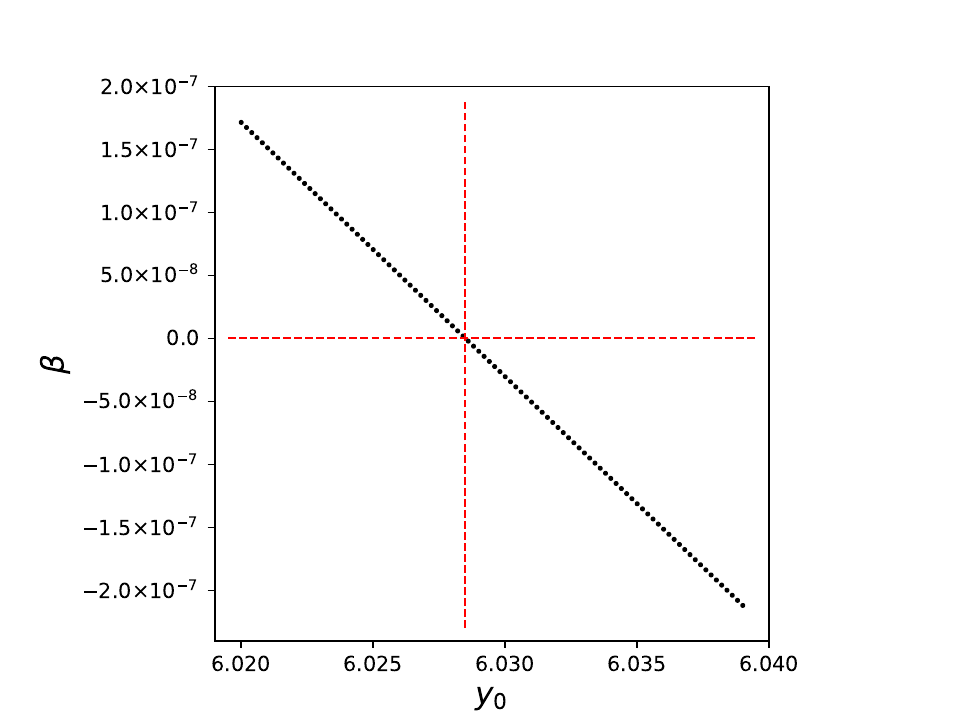}
    	\caption{}
    	\label{fig:beta_bubble}
    \end{subfigure}
    \begin{subfigure}{0.32\textwidth}
    	\includegraphics[width=\linewidth]{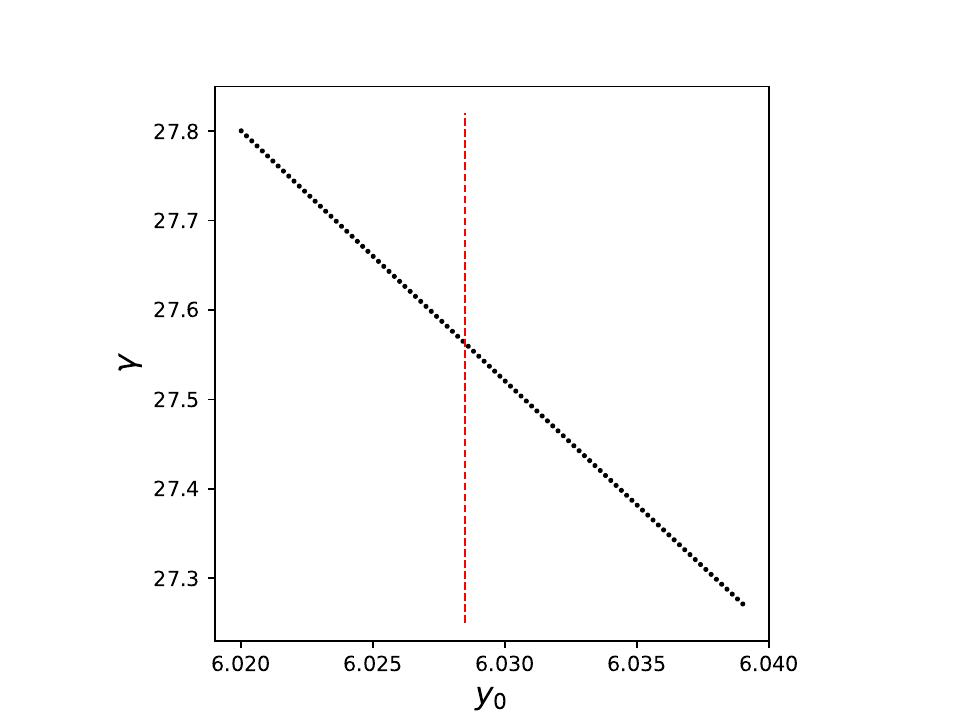}
    	\caption{}
    	\label{fig:gamma_bubble}
    \end{subfigure}
    \caption{
    The quantities $\Re \lambda_1, \Re \lambda_2, \beta, \gamma$ defined by \eqref{eq:betagamma} with $L$ replaced by $L_{\text{gsb}}$ are shown for the parameters studied in figure \ref{fig:angle_bubble_m1}
    ; i.e. for $\alpha= -1$, $\yp = 0$ with $\yo$ close to the transition at  $\yo \approx 6.028494$ (vertical dashed red line).
     {\bf (a):} To numerical precision $\Re \lambda_1$ (red) agrees with $\Re \lambda_2$ (blue).  {\bf (b):} At the transition (red vertical dashed line) we find $\beta=0$.  {\bf (c):} But $\gamma$ does not vanish at the transition, so $L_{\text{gsb}}$ is not diagonalizable.}
    \label{fig:bubblenotdiag}
\end{figure}

A short calculation gives
\begin{equation}
\label{eq:genbg}
\beta=\mathcal{N}^2(\lambda_+-\lambda_-) \ \ \ {\rm and}
\ \ \  \gamma=\mathcal{N}^{-2}(\lambda_+-\lambda_-)
  \end{equation}
  in terms of the eigenvalues $\lambda_\pm$ of $v_\pm$ at each $p$.
  Note that
equations \eqref{eq:v1eq} and \eqref{eq:gamma2} require $\lambda_1 = \lambda_2 = \lambda_\pm$ at the wall as well as $\beta =0$ (where the latter also follows from \eqref{eq:genbg}).  Plotting these quantities in figure \ref{fig:bubblenotdiag} shows that $\gamma$ does not vanish at the wall, so that our operators indeed fail to be diagonalizable.

\subsubsection{Rule-of-thumb contours near bubble walls}
\label{sec:contours}

Let us now turn to the implications for the rule-of-thumb contours.  As already noted, this contour will fail to be well-defined at the wall itself.  But it is interesting to compare the contours on opposite sides of the wall.  This can be done by again using the fact that, near the transition, the relevant pair of eigenvectors effectively evolves only in some fixed two-dimensional space (since they remain orthogonal to the infinite number of other eigenvectors which are continuous across the wall, and which thus evolve negligibly inside a region close enough to the wall).  We may thus focus only on how the Wick-rotation acts in this two-dimensional space as determined by the  associated negative- and positive-norm eigenvectors (or by the real and imaginary parts of the complex eigenvector).  Here it is important to note that, while the inner product on this space will be nearly constant when we are close to the wall, it will nevertheless vary at first order in the difference between $p$ and its value $p_*$ at the wall.  This is the same order at which the eigenvectors differ from each other.  As a result, if the positive-norm eigenvectors at some $p_1$ and $p_2$ happen to coincide, at linear order the orthogonality of positive- and negative-norm eigenvectors at both $p_1$ and $p_2$ would not require agreement between the negative-norm eigenvectors at $p_1$ and $p_2$.  In this sense, one should think of the positive- and negative-norm eigenvectors as being independent at this order, so that we must find both in order to determine the Wick-rotated contour.

In order to both confirm this picture and to parameterize the relevant two-dimensional space of perturbations, let us choose some fixed point $\tilde p$ in parameter space that lies close to the transition, but slightly to one side.  For example, in the present case we may choose some $\tilde y_0<y_0^*$ that is close to the edge of the bubble at $y_0^*$.  We denote the relevant two physical eigenvectors at general $p$ by $v_\pm$, and at the particular value $\tilde p$ by $\tilde v_\pm$.  Here $\pm$ indicates the sign of each eigenvector's norm (or the corresponding real and imaginary parts in the case of complex eigenvalue).

Since the fluctuation operator is diagonalizable at $\tilde p$, any vector can be written as a linear combination of the eigenvectors at $\tilde y_0$.  In particular, we may write
\begin{equation}
\label{eq:decompose_bubble}
\begin{split}
    v_-=A_+ \tilde v_+ + A_- \tilde v_-+v_\perp,\\
    v_+=B_+ \tilde v_+ + B_- \tilde v_-+u_\perp,
    \end{split}
\end{equation}
Here we have allowed for the fact that the two-dimensional space spanned by our $v_\pm$ does in fact vary slightly with $p$ by including the terms $v_\perp, u_\perp$ which are orthogonal to $\tilde v_\pm$.  Such $v_\perp, u_\perp$ clearly vanish at $p=\tilde p$, and will find numerically that they remain small over the range of $p$ we explore.  This then verifies that the space spanned by $v_\pm$ remains approximately constant over this range of parameters.

\begin{figure}[h!]
    \centering
    \begin{subfigure}{0.49\textwidth}
        \includegraphics[width=\linewidth]{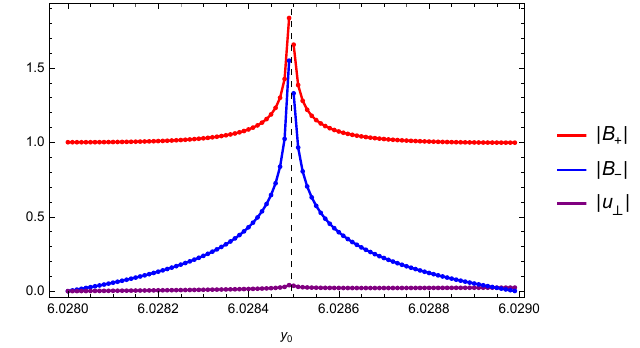}
        \caption{}
        \label{fig:bubble_real_G}
    \end{subfigure}
    \begin{subfigure}{0.49\textwidth}
        \includegraphics[width=\linewidth]{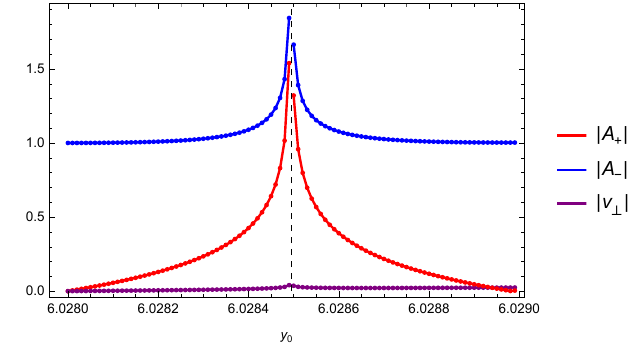}
        \caption{}
        \label{fig:bubble_imag_G}
    \end{subfigure}
    \begin{subfigure}{0.49\textwidth}
        \includegraphics[width=\linewidth]{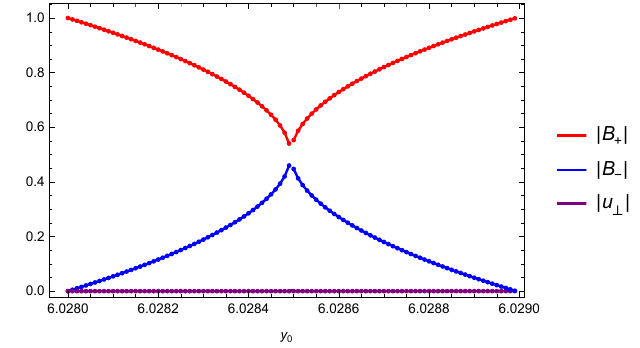}
        \caption{}
        \label{fig:bubble_real_E}
    \end{subfigure}
    \begin{subfigure}{0.49\textwidth}
        \includegraphics[width=\linewidth]{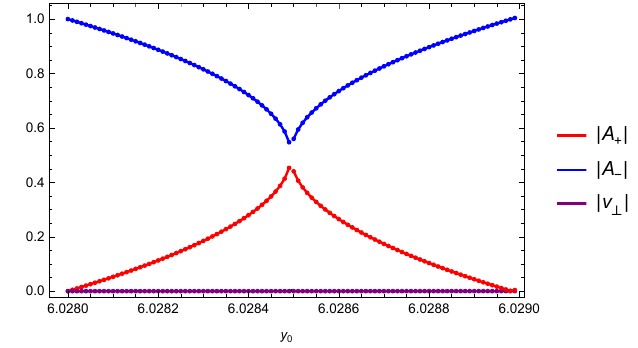}
        \caption{}
        \label{fig:bubble_imag_E}
    \end{subfigure}
    \caption{Absolute values of the coefficients $A_\pm, B_\pm$ defined in equation \eqref{eq:decompose_bubble} for the two relevant eigenvectors $v_\pm(\yo)$ in terms of fixed $\tilde v_\pm$ that are eigenvectors at $\tilde\yo=6.0280.$ We also include the magnitudes $|v_\perp|, |u_\perp|$ (where the magnitudes are defined using the fixed Cartesian metric \eqref{eq:CartesianMetric}) of the residual parts $v_\perp, u_\perp$ orthogonal to $\tilde v_\pm$ (with orthogonality defined by the DeWitt metric $\gma$).   The dashed black line denotes the (approximate) value $\yo^*$ corresponding to the bubble wall.  In the top row, the eigenvectors $v_\pm(\yo), \tilde v_\pm$ are normalized using the DeWitt metric at $\yo$.  The bottom row shows equivalent data but with the convention that $v_\pm,\tilde v_\pm$ are instead noramlized with respect to the fixed Cartesian metric \eqref{eq:CartesianMetric}.   While the value $\tilde \yo$ is at the extreme left edge of each plot, we see that the data is nearly symmetric about bubble wall (vertical dashed line) at $\yo^*$.  This symmetry improves as we take $\tilde \yo$ closer to $\yo^*$, supporting the idea that the contours on opposite sides of the wall are related by small smooth deformations.}
    \label{fig:comp_div_bubble}
\end{figure}

Numerical results for the parameters $A_\pm, B_\pm$ and for $u_\perp, v_\perp$ are shown in figure \ref{fig:comp_div_bubble}.
Without loss of generality, we normalize $\tilde v_\pm$ using our convention in equation \eqref{eq:vnorm}. But it is interesting to plot the results using two different normalizations for $v_\pm$.  In the top row of figure \ref{fig:comp_div_bubble}, we normalize each $v_\pm(p)$ using the DeWitt metric at $p$.  In this case, the the coefficients $A_\pm$ and $B_\pm$ must diverge at the value $p_*$ where $v_\pm$ becomes null.   (The magnitudes of $u_\perp, v_\perp$ diverge as well, but these pieces are small enough that the divergence is hard to see with the scale shown in the figure.)  In contrast, on the lower row we normalize both $\tilde v_\pm$ and each $v_\pm(p)$ using the fixed Cartesian metric \eqref{eq:CartesianMetric}.

The important feature of figure \ref{fig:comp_div_bubble} is that, while the reference parameter value $\tilde p$ appears on the left edge of each plot, the coefficients $A_\pm$, $B_\pm$ return to essentially the same values near the right edge of the plot.  In fact, the plots are nearly symmetric about the value $p_*$ corresponding to the bubble wall.  This approximate symmetry improves as we take $\tilde p$ closer to $p_*$.  We take this as evidence that the rule-of-thumb contours defined on opposite sides of the wall are essentially identical, and in particular are related by smooth deformations.  Recalling that our integrand is Gaussian, and thus that it can have no singularities at finite arguments, Cauchy's theorem (and the fact that all such contours give convergent integrals) then tells us that the rule-of-thumb defines equivalent path integrals on each side of the bubble wall.  In this sense, the fact that the rule-of-thumb fails at the bubble wall itself turns out to be harmless.

\subsubsection{Spectrum as a function of $\alpha$}
\label{sec:alphadep}

\begin{figure}[h!]
     \centering
     \begin{subfigure}{0.49\textwidth}        \includegraphics[width=\linewidth]{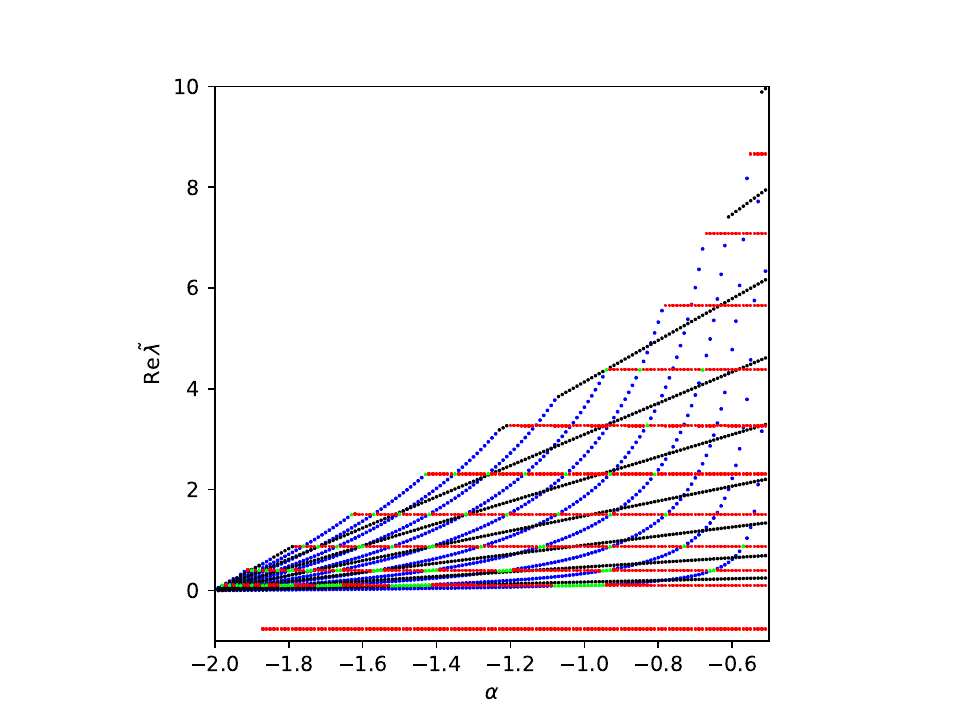}
         \caption{$\yp=0, \yo=10$}
     \end{subfigure}
     \begin{subfigure}{0.49\textwidth}
         \includegraphics[width=\linewidth]{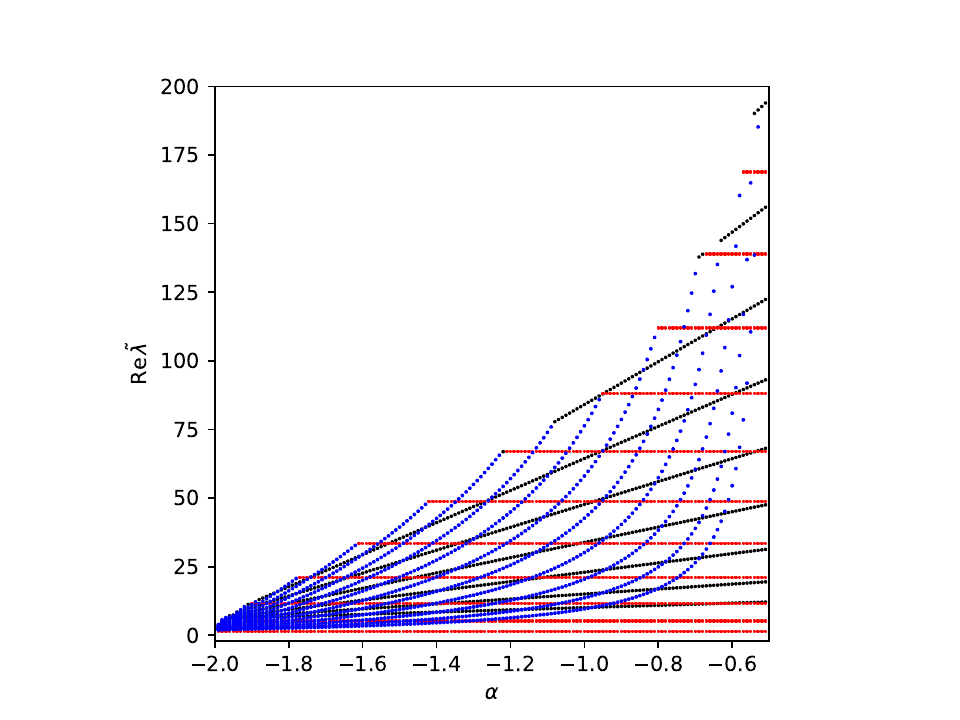}
         \caption{$\yp=1, \yo=10$}
     \end{subfigure}
        \caption{The real part of the lowest $21$ eigenvalues as a function of $\alpha$ for $\Lambda=0$ (left) and for a case with $\Lambda\neq 0$ (right) for $d=4$.  Real eigenvalues with positive (negative) norm are shown in red (blue), while the real part of a pair of complex-conjugate eigenvalues is shown in green.  Pure-gauge modes are shown in black and have positive norm. The case shown at left is thermodynamically unstable and correspondingly shows a single positive-norm mode with negative eigenvalue (depicted as an approximately horizontal line along the bottom of the plot). The case at right is thermodynamically stable.   Since these are the lowest $21$ modes at each value of $\alpha$, a given mode can appear or disappear from the plot at values of $\alpha$ where eigenvalues become degenerate (and in particular where the $21$st and $22$nd eigenvalues agree). }
        \label{fig:vary_alpha}
\end{figure}

Section \ref{sec:bubble} discussed the spectrum for each fixed $\alpha$ as a function of the choice of saddle.  But it is also interesting to examine the behavior of the spectrum for a fixed saddle as a function of $\alpha$.   Figure \ref{fig:vary_alpha} plots the lowest (in absolute value) $21$ modes as a function of $\alpha$ for two choices of $(\yo,\yp)$. The left panel describes a thermodynamically unstable black hole with $\yp=0$ and $\yo=10$ and correspondingly shows a single mode with negative eigenvalue\footnote{This mode disappears from the plot near $\alpha=-2$ due to the fact that, in that region, there are many modes with tiny eigenvalues and the plot includes only the 21 modes with smallest $|\tilde \lambda|$ at each $\alpha$. But the mode can be followed further with sufficient care.}, consistent with the results in figure \ref{fig:lowest_modes}. In contrast, the right panel shows a thermodynamically stable case for which all modes have positive eigenvalues. As expected from \eqref{eq:Geigenvalues}, the eigenvalues of the gauge modes are straight lines with a positive slope that intersect at $\tilde\lambda=6\yp^2$ when $\alpha=-2$.  Furthermore, as also expected from section \ref{sec:gaugeCondition}, the pure gauge modes all have positive norms for  $\alpha\in (-2,-1/2)$.

We see that the rule-of-thumb is consistent with thermodynamics for all $\alpha\in (-2,-1/2)$.
However, while the eigenvalues of the positive-norm physical modes are almost constant, the eigenvalues of the negative-norm physical modes decrease significantly as $\alpha$ decreases (though not in the strictly linear manner displayed by the pure gauge modes).  This suggests that new non-thermodynamic negative modes will arise for $\alpha < -2$, so that the rule-of-thumb will fail in that regime.  This phenomenon will be discussed and explained in  section \ref{sec:Gnotposdef} below.

\subsection{Failure of the rule-of-thumb for $\alpha<-2$}\label{sec:failure}
\label{sec:Gnotposdef}

We saw  in figure \ref{fig:vary_alpha} above that the eigenvalues of negative-norm physical modes decrease rapidly as $\alpha$ approaches $-2$.  In fact, we  observe numerically that all negative-norm physical modes have eigenvalues $\tilde\lambda=2\yp^2$ at $\alpha=-2$, though we have not found an analytic proof of this result.  Since the slope of the high modes is very steep at $\alpha=-2$, this suggests that additional negative modes will develop for any $\alpha < -2$, and thus that the rule-of-thumb will fail in that regime.

\begin{figure}[h!]
    \centering
    \begin{subfigure}{0.32\textwidth}
        \includegraphics[width=\linewidth]{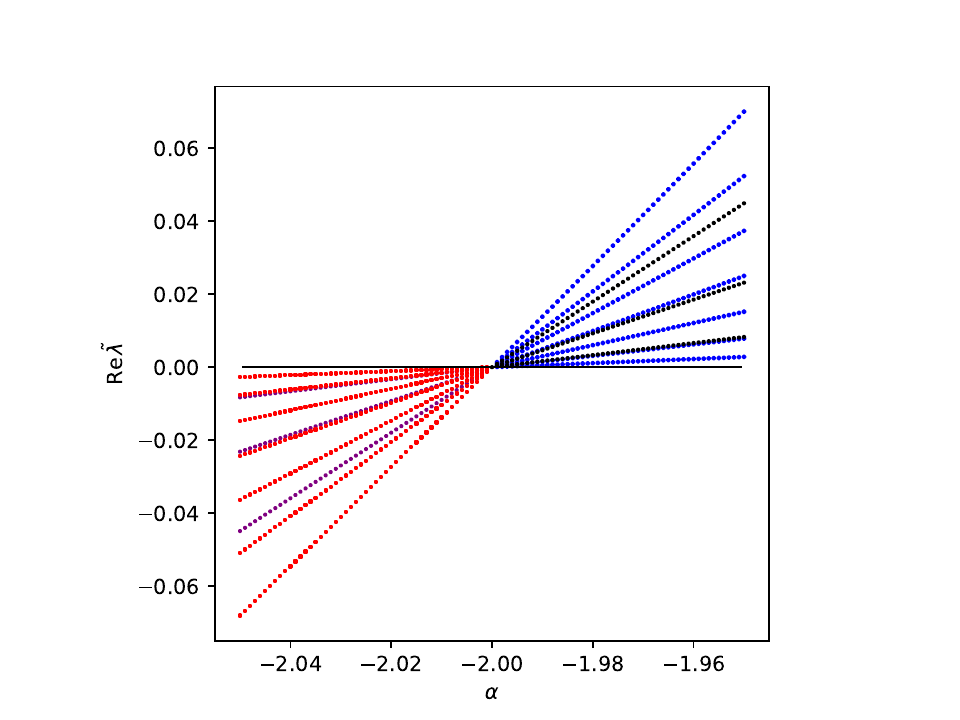}
        \caption{$\yp=0,\yo=10$}
        \label{fig:failFlat}
    \end{subfigure}
    \begin{subfigure}{0.32\textwidth}
        \includegraphics[width=\linewidth]{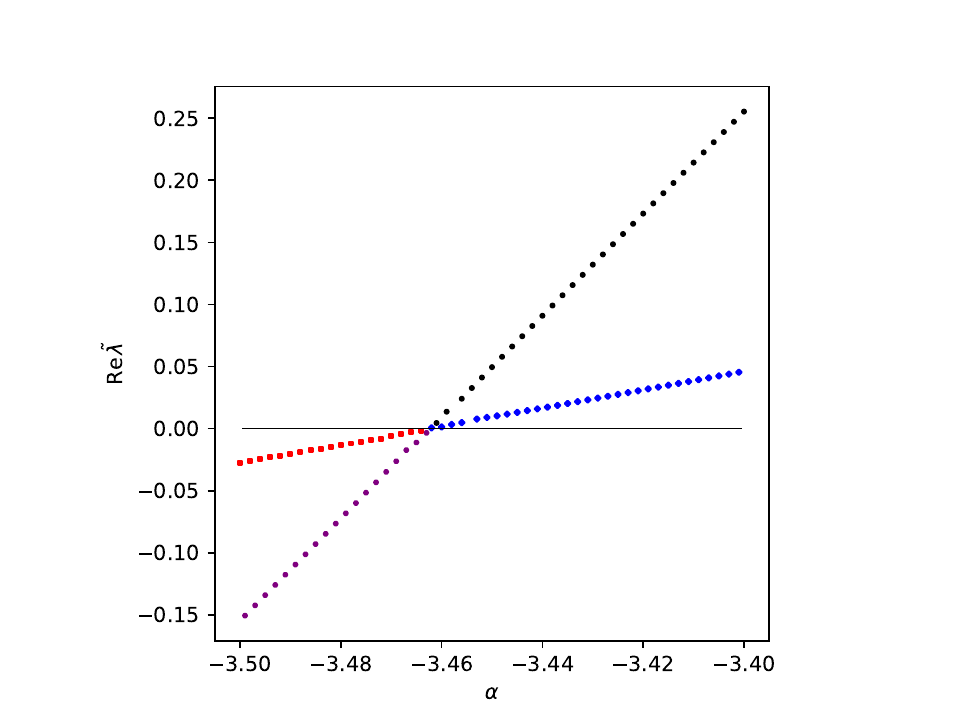}
        \caption{$\yp=1,\yo=10$}
        \label{fig:failAdS}
    \end{subfigure}
    \begin{subfigure}{0.32\textwidth}
        \includegraphics[width=\linewidth]{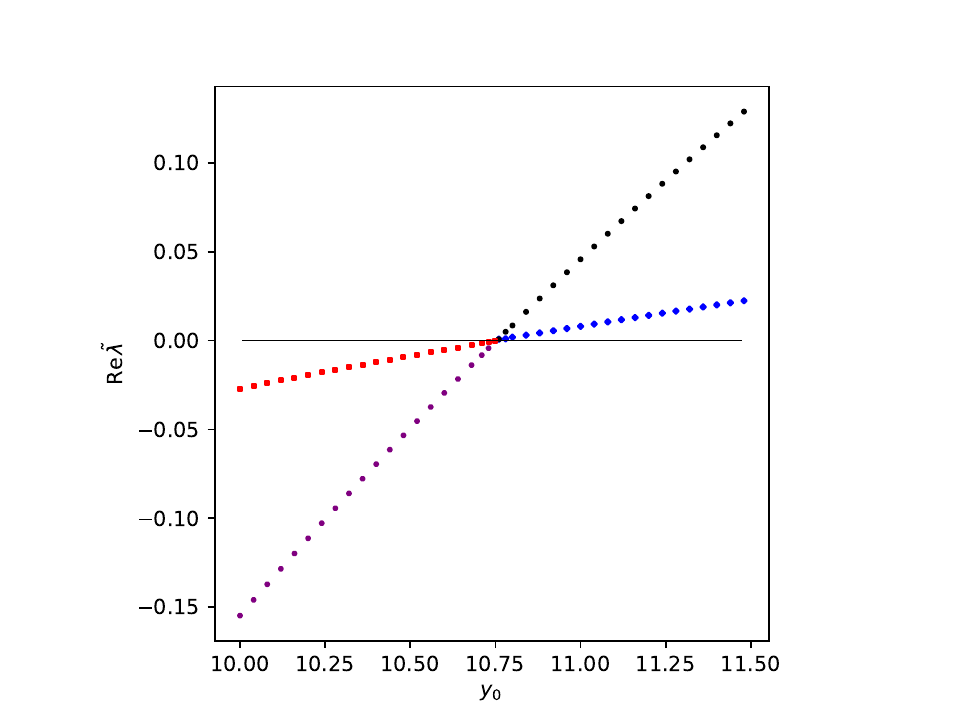}
        \caption{$\yp=1,\alpha=-7/2$}
        \label{fig:failAdS_small_alpha}
    \end{subfigure}
    \caption{ This is a continuation to smaller values of $\alpha$ of results shown previously in figure \ref{fig:vary_alpha}.  Again we set $d=4$.  As in \ref{fig:vary_alpha}, physical modes with positive (negative) norm are shown in red (blue) and pure gauge modes with positive norm are shown in black.  However, we now also find pure gauge modes with negative norm (purple).  The left panel plots the lowest 10 modes (as defined by the magnitude of the $\mathbb{L}$-eigenvalue) against $\alpha$ for the thermodynamically unstable case $\yp=0,\yo=10$. The middle panel
    considers the thermodynamically stable case $\yp=1,\yo=10$ and
    plots the last gauge mode to change sign as $\alpha$ decreases (i.e., the mode with the smallest value of $\lambda_{-1}$). For $\alpha<-2$, there are physical modes with negative eigenvalue and positive norm, though the particular physical mode shown crosses $\Re \tilde\lambda=0$ (black horizontal line) only at an $\alpha_*$ below $-3.46$. This shows explicitly that the rule-of-thumb fails for small enough $\alpha$. The right panel then shows a similar failure when we fix $\alpha = -7/2$ and instead vary $y_0$. All $y_0$ in this panel correspond to thermodynamically stable saddles. }
    \label{fig:failure}
\end{figure}

This failure will surprise some readers as it may appear to be excluded by the  discussion of section \ref{sec:Lops}. There we found that the vanishing eigenvalues of the fluctuation operators $L_\alpha$ coincide with having a vanishing eigenvalue of $L_{-1}$ for all $\alpha < -2/d$.  Nevertheless, as shown in figure \ref{fig:failAdS_small_alpha}, plots of our physical eigenvalues do indeed cross through $\tilde \lambda =0$ at certain $\alpha < -2/d$.

We address this puzzle by recalling that, since the metric $\gma$ is indefinite, self-adjoint operators need not be diagonalizable when their spectra degenerate.    The association with degenerate spectra can be seen by noting that (since any of our discretizations admits only a finite-dimensional space of perturbations), when the roots of the characteristic polynomial $\det(\mathbb{L} - \lambda \mathbb{1})$ are all distinct we may use such root to construct a complete basis of eigenvectors.    Thus $\mbl$ can fail to be diagonalizable at values of $\alpha$ where the eigenvalue of a positive-norm mode becomes degenerate with the eigenvalue of a negative-norm mode.

\begin{figure}[h!]
    \centering
    \includegraphics{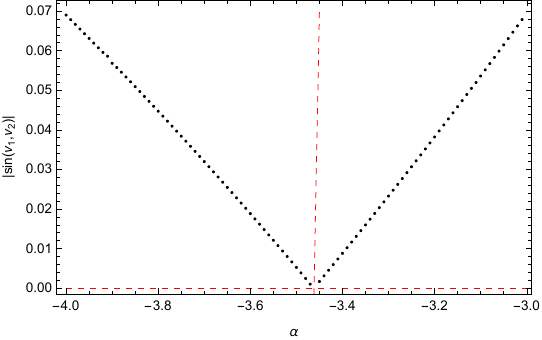}
    \caption{Absolute value of sine of the angle between the two relevant eigenvectors $v_{1,2}$ in figure \ref{fig:failAdS}. The horizontal red dashed line marks $|\sin (v_1,v_2)|=0$, i.e., the two vectors are linearly dependent. The vertical red dashed line marks the position of $\alpha_*$ where the eigenvalues of the two relevant modes vanish. We can see that as we approach $\alpha^*$, the two relevant eigenvectors become more and more linearly dependent.}
    \label{fig:check_diag}
\end{figure}

\begin{figure}[h!]
    \centering
    \begin{subfigure}{0.32\textwidth}
    	\includegraphics[width=\linewidth]{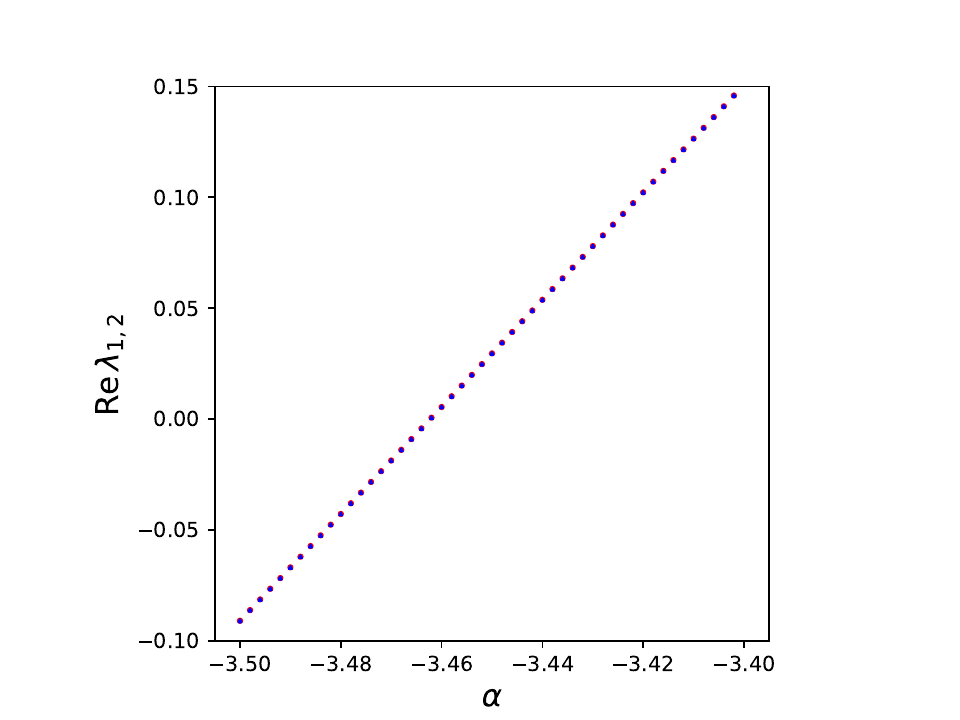}
    	\caption{}
    	\label{fig:lambda_small_alpha}
    \end{subfigure}
    \begin{subfigure}{0.32\textwidth}
    	\includegraphics[width=\linewidth]{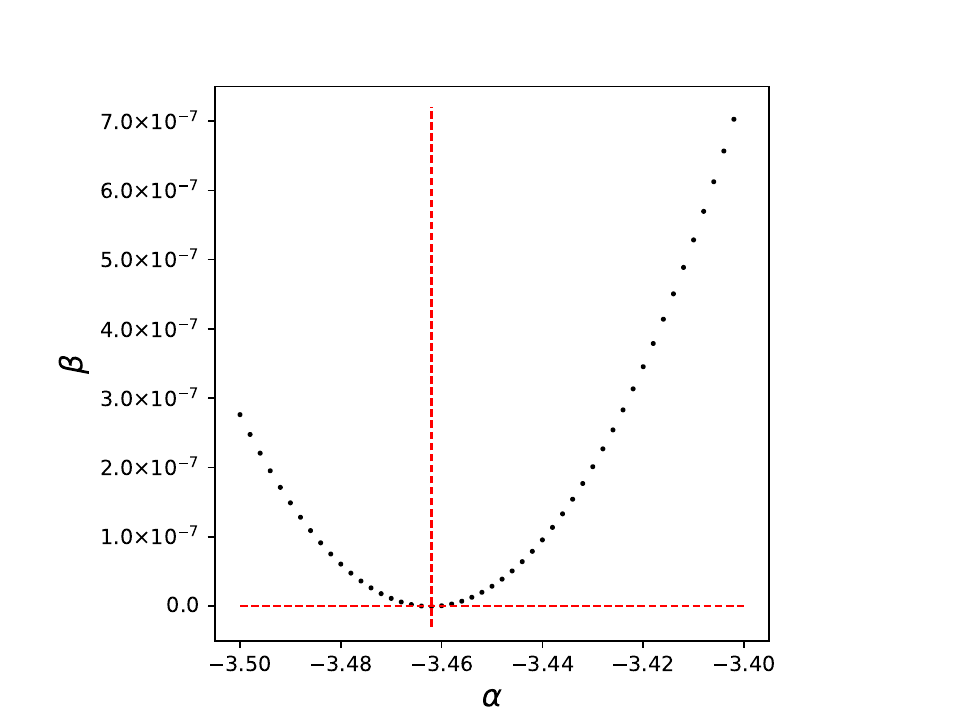}
    	\caption{}
    	\label{fig:beta_small_alpha}
    \end{subfigure}
    \begin{subfigure}{0.32\textwidth}
    	\includegraphics[width=\linewidth]{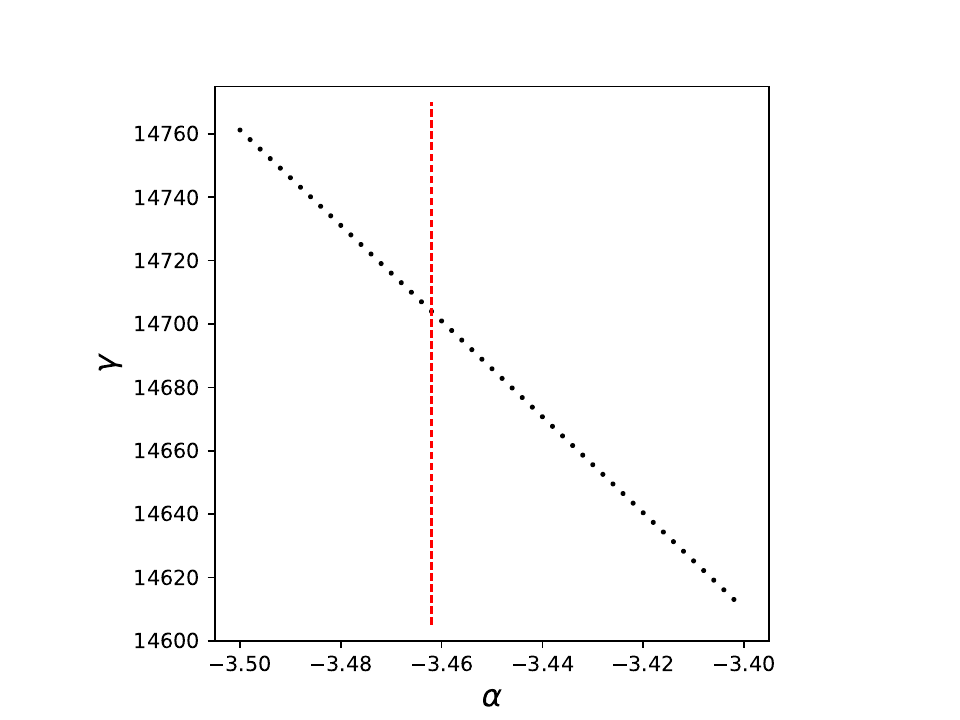}
    	\caption{}
    	\label{fig:gamma_small_alpha}
    \end{subfigure}
    \caption{ The quantities $\Re \lambda_1$, $\Re \lambda_2$, $\beta$, $\gamma$ defined by \eqref{eq:betagamma} with $L$ replaced by $L_{\text{gsb}}$ are shown for the parameters studied in figure \ref{fig:failAdS}; i.e. for $\yp= 1$, $\yo = 10$ with $\alpha$ in an interval close to the transition at  $\alpha \approx -3.462$ (vertical dashed red line).   As expected, we find that $\Re \lambda_1$ (red)  and $\Re \lambda_2$ (blue) agree to numerical precision (left panel).  The data for $\beta$ (middle pannel) is consistent with vanishing at the transition.  But $\gamma$ does not vanish at the transition (right panel),  so $L_{\text{gsb}}$ is not diagonalizable there.}
    \label{fig:smallalphanotdiag}
\end{figure}

We have already seen this sort of phenomenon at work at the bubble walls discussed in section \ref{sec:bubble}.  Figures \ref{fig:check_diag} and \ref{fig:smallalphanotdiag} below are the analogues for the current transition of the bubble wall figures \ref{fig:angle_bubble} and \ref{fig:bubblenotdiag}.  As in the bubble wall case, they verify that our eigenvectors coincide at the transition, and that $L_{\text{gsb}}$ fails to be diagonalizable when it occurs.

Nevertheless, the present transition differs in several ways from the transition at the bubble wall.  The first is that the current degeneracy always occurs at $\tilde \lambda=0$.  The second is that, as seen in figure \ref{fig:failAdS_small_alpha}, the current issue results from a physical mode becoming degenerate with a pure-gauge mode.  These features are clearly related since any pure-gauge mode is annihilated by $L$, and is thus a zero-eigenvalue eigenvector of this operator. However, since figure \ref{fig:failAdS_small_alpha} shows eigenvalues of $L_{\text{gsb}}$, we see that the operator $K = L_{\text{gsb}}-L$ also develops a zero-eigenvalue at this transition.  Thus by the discussion of section \ref{sec:gaugeCondition}, this can happen only when our gauge condition fails.

We now pause to explain the connections observed above.  In particular, we will show that $L$ fails to be diagonalizable due to mixing between a gauge and non-gauge mode at some value of parameters $p = (\yp,\yo,\alpha)$ {\it if and only if} $G$ is degenerate at $p$.

Let us begin with the only if direction.  The failure of $L$ to be diagonalizable in this way means that, at such $p$, there is a mode $h$ that is not pure-gauge but which maps to a pure-gauge mode $Lh$ under the action of $L$.  However, recall that for non-degenerate $G$ we may write the entire space of perturbations as $W^\perp \oplus W$.  Since $L$ annihilates $W$, the perturbations of the form $Lh$ can in fact be obtained using just those $h\in W^\perp$.  But for such $h$ we have $L_{\text{gsb}} h = Lh$.  The observation in the paragraph above then implies that, if $L$ fails to be diagonalizable due to degeneracy with a pure-gauge mode, when $G$ is invertible we would also have to find that $L_{\text{gsb}}$ also fails to be diagonalizable due to degeneracy with a pure-gauge mode.  The same would clearly be true if we replace  $L_{\text{gsb}} = L+K$ with $L+\eta K$ for $\eta\neq 1$.  But then the eigenvalues of the pure gauge mode would be proportional to $\eta$ (with non-zero coefficients), while the eigenvalues of the physical modes would remain fixed.  This would lift any degeneracies between physical and pure-gauge modes and forbid the required behavior.  It follows that failures of diagonalizability due to degeneracies with pure-gauge modes can occur only when $G$ fails to be invertible.

Recall from section \ref{sec:gaugeCondition} that $G$ can indeed fail to be invertible for $\alpha \le -2$.    Since we work in a cavity, the spectrum of $G$ will be discrete.  As a result,  invertibility will fail only for a measure-zero set of parameters $p= (\alpha,\yo,\yp)$ at which individual eigenvalues cross zero.  For other values of $p$ we may continue to use our chosen gauge-fixing scheme and to classify the modes as physical and pure-gauge in the manner described above.

We now turn to the if direction, showing that a zero eigenvalue for $G$ in fact requires the above non-diagonalizability of $L$ and $L_{\text{gsb}}$.  To begin, let us recall  equation \eqref{eq:signlambdanorm} which states that the sign of an eigenvalue of $G$ will always agree with the sign of the norm of the corresponding eigenvector, and that the norm of an eigenvector must vanish when its eigenvalue if zero.  In particular, when a pure-gauge eigenvalue changes from positive to negative, by \eqref{eq:signlambdanorm} the corresponding pure gauge eigenvector will also transition from positive to negative norm.    The correlation with the behavior of the physical modes may thus be explained by noting that, at a given level of discretization, any DeWitt$_\alpha$ inner product will have a definite signature involving some number $n_+$ ($n_-$) of positive (negative) signs.  Furthermore, since the inner product depends continuously on $\alpha$, the numbers $n_\pm$ can change only if the metric becomes degenerate. This occurs only at $\alpha=-2/d = -1/2$ (since here $d=4$), which is far from the range $\alpha < -2$ considered here.     It then follows that at some $\alpha_*$ a gauge-mode can change from being positive-norm to negative-norm only if there is some other mode that, at the same value of $\alpha_*$ changes from being negative-norm to positive-norm.  But \eqref{eq:Geigenvalues} implies that, in our framework, this second mode cannot be pure gauge.  It must thus be a negative-norm physical mode for $\alpha > \alpha_*$ that becomes a positive-norm physical mode below $\alpha_*$. The requirement of degeneracy with the pure-gauge mode then forces the physical mode eigenvalue to pass through $\lambda=0$ at the transition.  The fact that $L$ is non-diagonalizable then follows as noted above from \eqref{eq:LalphaTr} which forbids new zero-eigenvalue eigenvectors from appearing at any $\alpha < -2/d$.

We have seen that in many ways the phenomenon shown in figure \ref{fig:failAdS_small_alpha} is quite similar to what occurs at the edges of any complex bubble.  However, the implications for our contour rotation scheme are very different.  Section \ref{sec:bubble} argued rule-of-thumb contours on opposite sides of bubble walls to be related by continuous deformations, and thus to define the same path integral.  But this is clearly not so in the current case since the path integral converges only on one side of the transition!
\begin{figure}[h!]
    \centering
    \begin{subfigure}{0.49\textwidth}
    	\includegraphics[width=\linewidth]{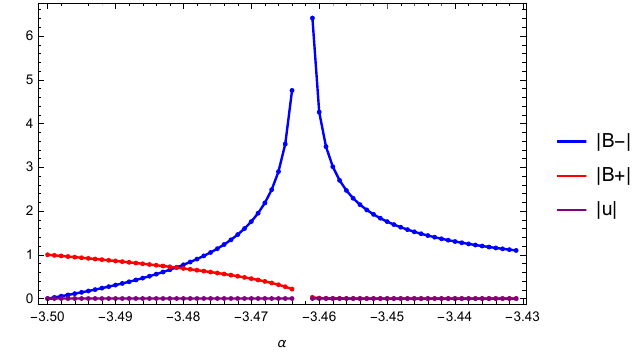}
    	\caption{}
    	\label{fig:alpha_p_G}
    \end{subfigure}
    \begin{subfigure}{0.49\textwidth}
    	\includegraphics[width=\linewidth]{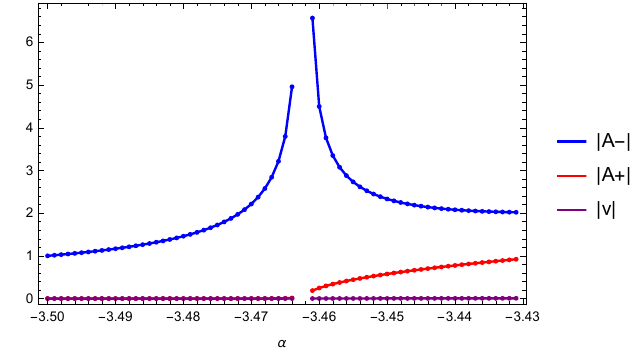}
    	\caption{}
    	\label{fig:alpha_n_G}
    \end{subfigure}
    \begin{subfigure}{0.49\textwidth}
        \includegraphics[width=\linewidth]{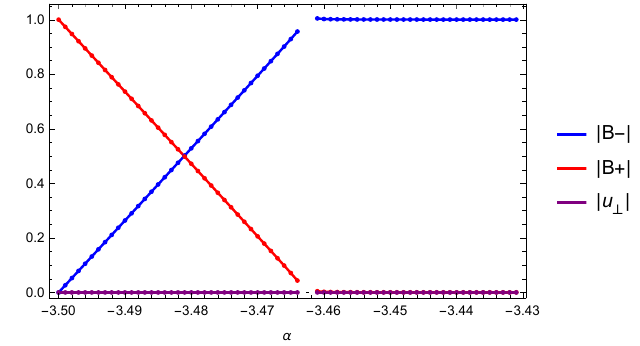}
        \caption{}
        \label{fig:alpha_p_E}
    \end{subfigure}
    \begin{subfigure}{0.49\textwidth}
        \includegraphics[width=\linewidth]{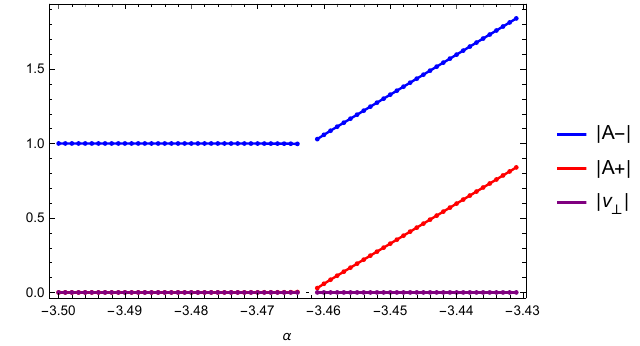}
        \caption{}
        \label{fig:alpha_n_E}
    \end{subfigure}

    \caption{
    Absolute values of the coefficients $A_\pm, B_\pm$ defined in equation \eqref{eq:decompose_bubble} for the two relevant eigenvectors $v_\pm$ at each $\alpha$ near the transition (at $\alpha \approx -3.462$) shown in the above figures.  These coefficients express $v_\pm$ in terms of fixed $\tilde v_\pm$ that are eigenvectors at $\alpha =-3.5$. We also include the magnitudes $|v_\perp|, |u_\perp|$ (where the magnitudes are defined using the fixed Cartesian metric \eqref{eq:CartesianMetric}) of the residual parts $v_\perp, u_\perp$ orthogonal to $\tilde v_\pm$ (with orthogonality defined by the DeWitt metric $\gma$).   The dashed black line denotes the (approximate)  position of the transition.  In the top row, the eigenvectors $v_\pm, \tilde v_\pm$ are normalized using the DeWitt metric at $\alpha$.  The bottom row shows equivalent data but with the convention that $v_\pm,\tilde v_\pm$ are instead noramlized with respect to the fixed Cartesian metric \eqref{eq:CartesianMetric}.
    The fact that $B_\pm$ are constant on one side of the transition while $A_\pm$ are constant on the other follows from the fact the same pure-gauge modes are eigenvectors for all $\alpha$ but that the norm of this mode changes sign at our transition.  Thus it represents the positive-norm eigenvector (shown at left) on one side of the transition while it represents the negative-norm eigenvector (shown at right) on the other.  The piecewise-linearity of the results is due to the fact that, as defined by the Euclidean metric used, the angle between the two perturbations is small (so that the perturbations are nearly co-linear). The plots are highly asymmetric, indicating that rule-of-thumb contours on opposite sides of the wall are markedly different.  This is consistent with the fact that the resulting path integral converges only on one side of the transition.}
    \label{fig:comp_alpha}
\end{figure}

Since Wick-rotations of pure-gauge modes have no effect, we may choose not to rotate them no matter what their eigenvalue may be for $L_{\text{gsb}}$.  With this convention, the difference between contours on the two sides is manifest as the number of physical directions Wick-rotated by the rule of thumb differs by one across the transition.  It is nevertheless reassuring to briefly consider the opposite convention in which we do indeed Wick-rotate pure-gauge modes according to the sign of their physical eigenvalue and to investigate the resulting contour rotations by constructing the analogue of figure \ref{fig:comp_div_bubble}.  The results (shown in figure \ref{fig:comp_alpha}) confirm that the rule-of-thumb contours are very different on opposite sides of the transition.

\section{Conclusion and Discussion}\label{sec:Conclusion}

Our work above studied possible contours for path integrals in linearized Euclidean Einstein-Hilbert quantum gravity with cosmological constant $\Lambda \le0 $.  In particular, we considered generalizations of the rule-of-thumb proposed in \cite{Marolf:2022ntb} associated with changing the parameter $\alpha$ in the DeWitt metric on the space of perturbations away from the value $\alpha=-1$ used in \cite{Marolf:2022ntb}.
Our work supports the idea that this generalized rule-of-thumb remains valid for $\alpha\in (-2,-2/d)$.  In particular, we returned to the study of the complex `bubbles' of eigenvalues identified in \cite{Marolf:2022ntb}. While found that them to be associated with breakdowns of the rule-of-thumb, we argued that the form of these breakdowns turns out to be harmless as the rule-of-thumb contours on opposite sides of the bubble wall were related by continuous deformations and define the same path integral.

Since the recipe requires a metric of indefinite signature, the fact that the DeWitt metric becomes positive definite for $\alpha > -2/d$ means that our generalized rule-of-thumb must fail in that regime.   We also showed it to fail for $\alpha\le -2$.  This issue is associated with the fact that induced metric on the space of pure-gauge modes is positive-definite only for $\alpha >-2$ (and for $\alpha=-2$ when $\Lambda < 0$).  In particular, we saw that the rule-of-thumb contour becomes ill-defined when the induced metric on the space of pure-gauge modes becomes degenerate, and that it is modified significantly when one crosses the codimension-1 barriers in parameter space associated with changing the signature of that induced metric.

\begin{figure}[h!]
	\centering
	\begin{subfigure}{0.49\textwidth}        \includegraphics[width=\linewidth]{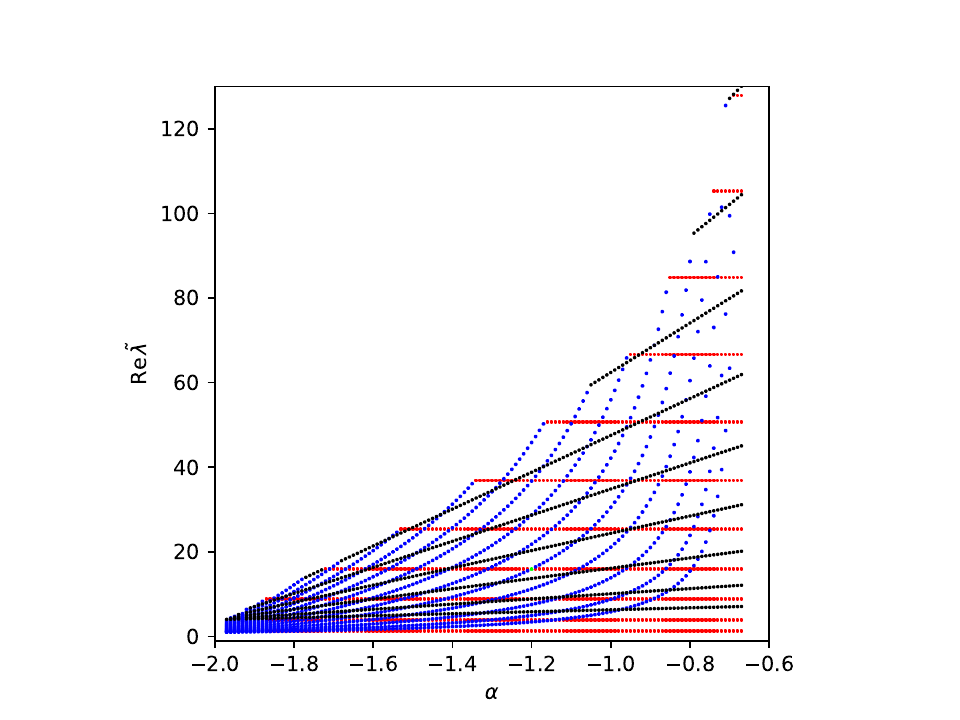}
		\caption{$d=3,\yp=1, \yo=10$}
	\end{subfigure}
	\begin{subfigure}{0.49\textwidth}
		\includegraphics[width=\linewidth]{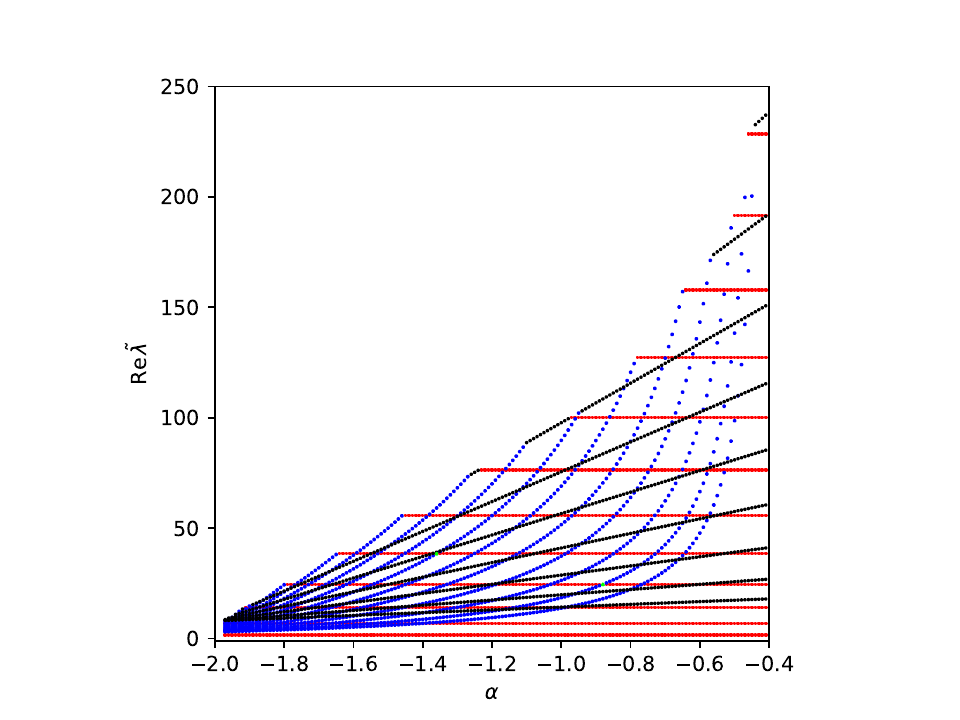}
		\caption{$d=5,\yp=1, \yo=10$}
	\end{subfigure}
	\caption{The real part of the lowest 21 modes as a function of $\alpha$ for AdS black holes in spacetime dimensions $d=3$ (left panel) and $d=5$ (right panel). The color coding is the same as that in figure \ref{fig:vary_alpha}. Both cases are thermodynamically stable and as expected, there is no negative eigenvalue. Note that the parameter $\alpha$ where the DeWitt metric become degenerate is $\alpha=-2/d$, where $d$ is the spacetime dimension. }
	\label{fig:other_dim_vary_alpha}
\end{figure}

While the numerical results used to motivate and illustrate these phenomena were all for spacetime dimension $d=4$, the final analytic arguments were valid for all $d \ge 3$. We thus expect similar results for all such cases. As confirmation, figure \ref{fig:other_dim_vary_alpha} shows the real part of the lowest 21 eigenvalues as a function of $\alpha$ for AdS-Schwarzschild (or BTZ) black holes with $\yp=1$ and $\yo=10$ for spacetime dimensions $d=3$ and $d=5$.

This suggests that, for general saddles of pure Einstein-Hilbert gravity with $\Lambda \le 0$, the rule-of-thumb can be applied using any DeWitt metric with $\alpha \in (-2, -2/d)$.  This hypothesis should be further tested by investigating other saddles, especially rotating saddles like AdS-Kerr.

The above result also admits a natural generalization to systems describing gravity coupled to matter.  For any such system, the space of perturbations will admit a class of ultra-local metrics in analogy with the DeWitt metrics.  In order to use the rule-of-thumb to solve the conformal-factor problem, one would need to choose such a metric with ``Lorentz-signature'' in the sense that it should have precisely one negative eigenvalue at each spacetime point.  Based on the above results, it is  then natural to conjecture that using the rule-of-thumb from \cite{Marolf:2022ntb} with this metric will give a valid contour when the induced metric on the space of pure-gauge perturbations is positive-definite.  It would be interesting to explore this hypothesis in both Einstein-scalar and Einstein-Maxwell systems.

It is clear that such studies only scrape the surface of the deeper question of what prescription might fundamentally define the contour of integration for Euclidean quantum gravity.  We may hope that it may some day be derived from a Lorentz-signature starting point, perhaps as in e.g. \cite{Marolf:2022ybi}, or from some more fundamental UV-complete theory.  But, until then, it may be useful to continue to take small steps toward exploring possible contours, including considering other boundary conditions (as in the microcanonical studies of \cite{Marolf:2022jra}), looking beyond the class of ultra-local metrics, studying examples with positive cosmological constant, and incorporating higher-derivative corrections.

\acknowledgments
XL and DM were supported by NSF grant PHY-2107939, and by funds from the University of California. J. E. S. has been partially supported by STFC consolidated grant ST/T000694/1.

\appendix

\section{Explicit Expression for the Modified Quadratic Action}\label{appendix: action}
This appendix provides the explicit form of the modified quadratic action defined by equation (\ref{eq:newaction}).  For the cases of interest here, the quadratic action may be written in the form
\begin{equation}
\begin{split}
   \check{S}^{(2)}[h]&=\frac{1}{32\pi G}\int_{\mathcal{M}}d^dx\sqrt{\hat{g}}(h_{ab}\hat{\mathcal{G}}^{ab cd}_{-1} [L_{-1}h)_{cd}+h_{ab}\hat{\mathcal{G}}^{ab cd}_{\alpha} \tilde{h}_{cd}]\\
   &=\frac{\Omega_{d-2}}{64\pi G}\int_{r_+}^{r_0}dr\, r^{d-2}\vec{q}\cdot\left[f\bm{Q}\cdot \frac{d^2\vec{q}}{dr^2}+\bm{P}\cdot\frac{d\vec{q}}{dr}+\bm{V}\cdot\vec{q}\right],
   \end{split}
\end{equation}
where $\vec{q}=\{a(r),b(r),c(r)\}$, $\bm{Q}$ and $\bm{V}$ are symmetric matrices while $\bm{P}$ is not necessarily symmetric.  For Einstein-Hilbert gravity linearized about ESAdS we find
\begin{equation}
    \begin{split}
        \bm{Q}_{11}=&-\alpha^2,\bm{Q}_{22}=-(2+\alpha)^2,\bm{Q}_{33}=-(d-2)(6-2\alpha^2+d(\alpha^2-2)),\\
        \bm{Q}_{12}=&-\alpha(2+\alpha),\bm{Q}_{13}=-(d-2)(\alpha^2-2),\bm{Q}_{23}=-(d-2)(\alpha+2)\alpha,\\
        \bm{V}_{11}=&\,\frac{f'^2}{f}+\frac{\alpha(d-2)f'}{r}+\alpha f'',\\
        \bm{V}_{22}=&-\frac{2(d-2)(d-5+(d-3)\alpha)f}{r^2}-\frac{(d-2)(4+3\alpha)f'}{r}+\frac{f'^2}{f}-(\alpha+2)f'',\\
        \bm{V}_{33}=&\,\frac{2(d-2)(d^2-7d+12)}{r^2}+\frac{2(d-2)^2(2+(d-3)\alpha)f}{r^2}+\frac{2(d-2)^2\alpha f'}{r},\\
        \bm{V}_{12}=&-\frac{(d-3)(d-2)(\alpha+1)f}{r^2}-\frac{(d-2)(\alpha+2)f'}{r}-\frac{f'^2}{f}+f'',\\
        \bm{V}_{13}=&\,\frac{(d-3)(d-2)}{r^2}+\frac{(d-3)(d-2)\alpha f}{r^2}+\frac{(d-2)(4+d\alpha)f'}{2r}+\frac{(d-2)\alpha f''}{2},\\
        \bm{V}_{23}=&\,\frac{(d-3)(d-2)}{r^2}-\frac{(d-2)(10+(d-5)d+\alpha(d-3)^2)f}{r^2}-\frac{(d-2)(4(d-3)-8\alpha+3d\alpha)f'}{2r}\\ & -\frac{(d-2)(\alpha+2)f''}{2},
    \end{split}
\end{equation}
and $\bm{P}$ is given by
\begin{equation}
    \begin{split}
        \bm{P}_{11}=&-\frac{(d-2)\alpha^2 f}{r}-\alpha^2 f',\\
        \bm{P}_{12}=&-\frac{(d-2)(2+\alpha(4+\alpha))f}{r}-(2+\alpha(4+\alpha))f',\\
        \bm{P}_{13}=&\,\bm{P}_{23}=-\frac{(d-2)(2-2d-2\alpha+(d-2)\alpha^2)f}{r}-(d-2)(\alpha^2+\alpha-1),\\
        \bm{P}_{21}=&-\frac{(d-2)(\alpha^2-2)f}{r}-(\alpha^2-2)f',\\
        \bm{P}_{22}=&-\frac{(d-2)(\alpha+2)^2f}{r}-(\alpha+2)^2 f',\\
        \bm{P}_{31}=&-\frac{(d-2)(2(3-d)+2\alpha+\alpha^2(d-2))f}{r}-(d-2)(\alpha^2-\alpha-3)f',\\
        \bm{P}_{32}=&-\frac{(d-2)(2(d-1)-2\alpha(3+\alpha)+d\alpha(4+\alpha))f}{r}-(d-2)(\alpha^2+3\alpha+1)f',\\
        \bm{P}_{33}=&-\frac{(d-2)^2(2(d-3)+\alpha^2(d-2))f}{r}-2(d-2)(3-d+\alpha^2(d-2))f'.\\
    \end{split}
\end{equation}
Here the primes denote derivatives with respect to the radial coordinate $r$, and $\Omega_{d-2}$ is the volume of the metric on a unit radius round $(d-2)$-sphere.

~~~~~~~~~~~~
\addcontentsline{toc}{section}{References}
\bibliographystyle{JHEP}
\bibliography{references}

\end{document}